\documentclass[usenatbib]{mn2e}
\usepackage{epsfig}
\usepackage{epstopdf}
\usepackage{graphicx}
\usepackage{subfigure}
\usepackage{amssymb}
\usepackage{amsmath}
\usepackage{color}
\usepackage{float}

\newcommand{\bn}{\begin{enumerate}}
\newcommand{\en}{\end{enumerate}}
\newcommand{\bi}{\begin{itemize}}
\newcommand{\ei}{\end{itemize}}

\def\gtorder{\mathrel{\raise.3ex\hbox{$>$}\mkern-14mu
    \lower0.6ex\hbox{$\sim$}}}
\def\ltorder{\mathrel{\raise.3ex\hbox{$<$}\mkern-14mu
    \lower0.6ex\hbox{$\sim$}}}

\newcommand{\apj}{ApJ}

\newcommand{\apjl}{ApJL}
\newcommand{\apjs}{ApJS}
\newcommand{\mnras}{MNRAS}
\newcommand{\aj}{AJ}
\newcommand{\araa}{ARA\&A}

\newcommand{\pasj}{PASJ}

\newcommand{\prd}{Phys.~Rev.~D}

\title[Direct Collapse to Supermassive Black Hole Seeds with Radiative Transfer: Isolated Halos]
{Direct Collapse to Supermassive Black Hole Seeds with Radiative Transfer: Isolated Halos}

\author[Luo et al.]
{Yang Luo$^{1,2}$\thanks{E-mail: yluo@uky.edu},
Kazem Ardaneh$^{1}$, Isaac Shlosman$^{1,2}$\thanks{E-mail: shlosman@pa.uky.edu}, Kentaro Nagamine$^{1,3}$, \newauthor
John H. Wise$^{4}$, Mitchell C. Begelman$^{5}$
\\
$^{1}$ Theoretical Astrophysics, Department of Earth \& Space Science, Graduate School of Science,
  Osaka University, Osaka 560-0043, Japan\\
$^{2}$ Department of Physics \& Astronomy, University of Kentucky, Lexington, KY 40506-0055, USA\\
$^{3}$ Department of Physics \& Astronomy, University of Nevada, Las Vegas, NV 89154-4002, USA\\
$^{4}$ Center for Relativistic Astrophysics, Georgia Institute of Technology, Atlanta, GA 30332-0430, USA\\
$^{5}$ JILA, University of Colorado \& NIST, 440 UCB, Boulder, CO 80309-0440, USA\\
}
  
\begin{document}

\date{Accepted: February 1, 2018; Received: February 1, 2018; in original form: December 19, 2017}


\maketitle

\begin{abstract}
Direct collapse within dark matter (DM) halos is a promising path to form supermassive black hole (SMBH)
seeds at high redshifts. The outer part of this collapse remains optically thin, and has been studied
intensively using numerical simulations. However, the innermost region of the collapse is expected
to become optically thick and requires to follow the radiation field in order to understand its subsequent
evolution. So far, the adiabatic approximation has been used exclusively for this purpose. We apply
radiative transfer in the flux-limited diffusion (FLD) approximation to solve the evolution of coupled
gas and radiation, for isolated halos. 
For direct collapse within isolated DM halos, we find that (1) the photosphere forms at 
$\sim 10^{-6}$\,pc and rapidly expands outwards. (2) A central core forms, with a mass of $\sim 1\,M_\odot$,
supported by gas pressure gradients and rotation. (3) Growing gas and radiation pressure 
gradients dissolve it. 
(4) This process is associated with a strong anisotropic outflow, and another core forms nearby and grows 
rapidly. (5) Typical radiation luminosity emerging from the photosphere encompassing these cores is $\sim 5\times 
10^{37}-5\times 10^{38}\,{\rm erg\,s^{-1}}$, of the order the Eddington luminosity. (6) Two variability 
timescales are associated with this process: a long one, 
which is related to the accretion flow within the central $\sim 10^{-4}-10^{-3}$\,pc, and $\sim 0.1$\,yr,
which is related to radiation diffusion. (7) Adiabatic models have been run for comparison and their 
evolution differs profoundly from that of the FLD models, by forming a central geometrically-thick disk.
Overall, an adiabatic equation of state is not a good approximation to the advanced stage of direct collapse,
because the radiation is capable of escaping due to anisotropy in the optical depth
and associated gradients.
\end{abstract}


\section{Introduction}
\label{sec:intro}

A growing number of quasars found at redshifts $z\gtorder 6$, including one at $z\sim 7.54$ 
\citep{vene17,bana18},when the universe was younger than a Gigayear,
requires a very efficient way of forming early supermassive black holes (SMBHs) \citep[e.g.,][]{fan03,will10,
mor11,wu15}. While small black holes can form just after the Big Bang \citep[e.g.,][]{carr10}, SMBHs must 
wait until the gas can collapse within dark matter (DM) halos. SMBH seeds can form as the end products of 
stellar evolution, namely, of metal-free Population\,III stars 
\citep[e.g.,][]{hai01,abel02,bromm04,vol06,li07,pel07}, supermassive stars (SMS)
\citep[e.g.,][]{hae93,bromm03,beg06,wise08,beg09,mil09,reg09,sch10,hos11,choi13,choi15,lat13a,lat13b,shlo16}, 
and stellar clusters, either relativistic \citep[e.g.,][]{ipser69} or gas-rich \citep[e.g.,][]{deve09,lupi14}. 
In principle, it is also posssible that the stellar evolution stage can be by-passed completely, for example
if the gas never gets hot enough to ignite thermonuclear reactions \citep[e.g.,][]{beg09,choi13,shlo16}. 

In this work we focus on direct collapse scenarios, in which gas accumulates and collapses to form a SMBH seed
either with or without the intermediate stage of an SMS. Such models are often glossed over an important stage 
in the collapse, when it becomes optically thick, substituting an adiabatic approximation for 
a detailed study of the radiation hydrodynamics.  

Direct collapse can happen only when the virial temperature of DM halos exceeds the gas temperature.  
If the gas with a primordial composition is capable of forming
molecular hydrogen, halos with virial temperatures of $\sim 100-1000$\,K can suffice. In this case, 
gravitational collapse leads to one or a few Pop\,III stars per halo for $z\ltorder 50$, with an IMF  
initially thought to be top-heavy, $\sim 100-1,000\,M_\odot$ \citep[e.g.,][]{abel02,
bromm02,oshea07,bromm13}. Inclusion of radiative feedback indicates a rather normal IMF 
\citep[e.g.,][]{hos11,hir15,hos16}. Supersonic streaming velocities remaining from recombination can 
suppress formation of Pop\,III stars, allowing the gas to form a more massive central object \citep{hir17}.

If, however, the Pop\,III stars dissociate H$_2$ or prevent its formation altogether, collapse will be
triggered only for virial temperatures $T\gtorder {\rm few}\times 10^3$\,K. Suitable halos have masses of 
$10^7-10^8\,M_\odot$ and become abundant at $z\ltorder 20$.  
Under these latter conditions, it has been conjectured that direct collapse will lead to an SMS 
with a mass in the range of $\sim 10^4-10^6\,M_\odot$, if fragmentation can be suppressed and the angular 
momentum can be efficiently transferred out. 

\citet{beg09} argued that gravitational torques transfer the angular momentum in the collapsing gas
to the DM and the outer gas, which has been verified explicitly, both for isolated collapse and collapse
in a cosmological framework \citep{choi13,choi15}. Furthermore, they found that global instabilities in the
rotating collapsing gas lead to supersonic turbulent motions that damp fragmentation in the atomic gas.
Contrary to self-similar analysis, which was necessarily limited to a linear stage \citep{hana00},
the growing bar-like $m=2$ mode in its nonlinear stage did not lead to fragmentation, but induced gas inflow.
In all cases, the collapse is dominated by filamentary structure \citep[e.g.,][]{shlo16,luo16}.

The final stages of the collapse are expected to be characterized by radiation trapping, initially partial
and thereafter complete. Simple logic points to the formation of a central object, but its nature is
elusive. Is the collapse stopped early, leading to the formation of a hydrostatic object, an SMS, whose
subsequent evolution leads to the formation of the SMBH seed? Or can the collapse 
proceed directly to an SMBH seed?

If the SMS forms, it has been conjectured that the followup nuclear burning and
core collapse leave an SMBH seed of $\sim 10-10^5\,M_\odot$, which grows rapidly via hypercritical accretion.
Such a pre-collapse object has the structure of a hylotrope, and the post-collapse configuration has been 
termed a quasistar  \citep[e.g.,][]{beg06,beg08,beg10}. 

The formation details of the SMS, however, appear to be murky. When does the photosphere form and where, what 
is its shape, and what is the 
effective temperature? How does an SMS get rid of the angular momentum in the 
collapsing gas? Does it rotate as a star, i.e., with surfaces of constant angular momentum, $J$, that resemble
ellipsoids of rotation? Or does it rotate as a disk, i.e., with iso-$J$ surfaces of a cylindrical shape?
Are the central conditions
sufficient to trigger thermonuclear reactions? How efficient is convection? Is the formation of the SMS 
associated with radiation- or gas pressure-driven outflows? 

Models of the optically-thin part of the collapse within DM halos, on scales of $\sim 1$\,kpc down to 
$\sim 1$\,AU, have emphasized various aspects of this stage: from formation and effects of molecular
hydrogen, to Lyman\,$\alpha$ diffusion, to background UV flux produced by Pop\,III stars, as we have 
referenced above.
However, inside $\sim 1$\,AU, the radiation pressure is expected to build up, and have both dynamical and
thermodynamical effects. Current modeling assumes that the radiation pressure buildup  within the optically
thick flow will lead it to follow an adiabatic equation of state 
\citep[e.g.,][]{bec15, bec17}. 

In this work, we test this assumption by treating the optically-thick part of the accretion flow using 
radiative transfer in the
Flux-Limited Diffusion (FLD) approximation. We follow the flow as the radiation pressure builds up
and becomes as important as the gas thermal pressure. Moreover, we evolve the adiabatic models to 
compare and contrast with the model involving radiative transfer.
In the current paper, we deal with an isolated DM halo, while in the accompanying paper \citep{arda18}, 
we invoke DM halos within a fully cosmological framework.

This paper is structured as follows. The next section describes the numerical aspects of our modeling, 
the details of radiation transfer solver implemented here, and the initial conditions used.
Sections\,3 and 4 present our results for adiabatic and non-adiabatic flows, respectively, and section\,5 
compares them. The last section summarizes our main conclusions from this work. We provide test models for 
radiative transfer in the Appendix. In the following, we abbreviate spherical radii with $R$ and 
cylindrical ones with $r$. 

\section{Numerical techniques}
\label{sec:method}

We use the modified version of the Eulerian adaptive mesh refinement (AMR) code Enzo-2.4 
\citep{bryan97,nor99}. Our modifications are explained in this section. 

Enzo uses a multigrid particle mesh $N$-body method to calculate 
gravitational dynamics including collisionless DM particles, and a second-order piecewise parabolic method 
\citep[][]{cole84,bryan95} to solve hydrodynamics. The structured AMR used in
Enzo allows additional inner meshes as the simulation advances to enhance the resolution in the user-desired 
region. It places no fundamental restrictions on the number of
rectangular grids used to cover some region of space at a given level of refinement, or on the number of levels of
refinement \citep{ber89}. A region of the simulation grid is refined by a factor of two in length scale, if
the gas or DM  densities become greater than $\rho_0N^l$, where $\rho_0$ is the minimum density above which 
refinement occurs, $N = 2$ is the refinement factor and $l$ is the maximal AMR refinement level. 

We use a maximal refinement level of 33, which corresponds to $10^{-8}$\,pc, although the code only reaches 
refinement level of 30, i.e., $8\times 10^{-8}$\,pc.

To avoid spurious fragmentation, we satisfy the \citet{true97} requirement for resolution of the Jeans
length, i.e., at least four cells per Jeans length. In fact, following recent numerical experiments, higher
resolution is required to properly resolve the turbulent motions \citep[e.g.,][]{sur10,fede11,turk12, 
lat13a}. Consequently, we have resolved the Jeans length with at least 16 cells.

\subsection{Radiation Hydro and Radiative Transfer}
\label{sec:rad-transf}

Radiation transport is modeled via the flux-limited diffusion (FLD) approximation.
In regions that are optically thick, in the sense of a ``true'' absorption modified by electron scattering, 
we assume local thermodynamic equilibrium (LTE), in which emissivity
is given by the Planck intensity, and gas ionization is determined by the Saha equation 
\citep[e.g.,][]{ryb79}. The radiative transfer is fully anisotropic, i.e., each grid cell is a source and
sink of radiation, communicates
with 6 neighboring cells, and the optical depth is calculated accordingly (section\,\ref{sec:cool-heat}). 

The resulting radiation
transport equation is solved using a fully implicit inexact Newton method. This solver, which
couples to the AMR cosmological hydro solver by an explicit, operator-split algorithm only at the end
of the top level timestep \citep{rey09}, has been modified
by us to update each refinement level at the end of its corresponding timestep, making the FLD fully
consistent with the hydro part.

We have modified the equations of \citet{rey09} by introducing the radiation force and $v/c$ order terms,
where $c$ is the speed of light and $v$ is the gas velocity.
The new Euler equation is 

\begin{equation}
\label{eq:euler}
\frac{\partial \rho \mathbf{v}}{\partial t}  
    + \nabla\cdot (\rho \mathbf{v} \mathbf{v} + \mathbb{I} p )  = 
    - \rho \nabla \phi + \frac{\kappa_{\rm R}}{c}{\bf F},
\end{equation}
where $\rho$ and $p$  are the baryon density and  thermal pressure, respectively. 
The matrix $\mathbb{I}$ is the identity matrix. 
The gravitational potential $\phi$ is calculated from the baryon density $\rho$ and DM density 
$\rho_{\rm DM}$. 
Here ${\bf F}$ is the radiation energy flux, and $\kappa_{\rm R}$ is the Rosseland mean opacity 
(\S\,\ref{sec:opacity}). Thus, the self-gravity of the gas is fully accounted for.

Under the FLD approximation, the radiative flux vector can be written in the form of Fick's diffusion law, 
i.e., is proportional to the gradient of radiation energy density \citep{lev81,lev84},

\begin{equation}
\label{eq:fld_approx}
 {\bf F} = - \frac{c\lambda}{\kappa_{\rm R}}\nabla E,
\end{equation}
where $\lambda = \lambda (E, \nabla E, \kappa_{\rm R}) = (9 + R^2)^{-1/2}$ is the {\it flux limiter}, 
$R={|\nabla E|}/{(\kappa_{\rm R} E)}$. Note that velocities encountered in our simulations are substantially
sub-relativistic, which allows to use this simple closure. The evolution of the
radiation energy density, $E$, is given by \citep{rey09,ENZO14},

\begin{equation}
\begin{aligned}\label{eq:radiation_PDE}
  \partial_{\rm t} E + &\nabla\cdot (E {\bf v}) = \\
  & - \nabla\cdot {\bf F} - \mathbb{P}:\nabla {\bf v}- c\kappa_{\rm P} E 
    + \eta  - \frac{\kappa_{\rm R}}{c}{\bf F}\cdot \mathbf{v},
\end{aligned}    
\end{equation}
where we have added the last term. Here $\mathbb{P}$ is the radiation pressure tensor written
with auxiliary functions,

\begin{equation}
\begin{aligned}
\label{eq:prad}
  &\mathbb{P} = \mathbb{D}E\\ 
  &\mathbb{D}=\frac{1-\chi}{2}\mathbb{I}+\frac{3\chi-1}{2}\mathbf{n}\otimes \mathbf{n}\\ 
  &\chi=\lambda+\lambda^2R^2\\
  &\mathbf{n}=\frac{\nabla E}{|\nabla E|}.
\end{aligned}
\end{equation}

The coefficients $\kappa_{\rm P}$ and $\kappa_{\rm R}$ are Planck   
and Rosseland mean opacities, respectively (\S\,\ref{sec:opacity}). The parameter $\eta$ is the blackbody 
emissivity given by $\eta = 4\kappa_{\rm P}\,\sigma_{\rm SB}\,T^4$, where $\sigma_{\rm SB}$ is the 
Stefan-Boltzmann constant and $T$ is the gas temperature.
The frequency-dependence of the radiation energy is omitted by integration over the radiation
energy spectrum.  

The equation for the evolution of the gas energy density $e$ has been modified as well by introducing the 
$v/c$ order term,

\begin{equation}
\begin{aligned}
 \frac{\partial e}{\partial t} +  & \nabla\cdot [(e+p)\mathbf{v}]  = \\
    & - \rho\mathbf{v}\cdot\nabla\phi 
    + c \kappa_{\rm P} E - \eta + \frac{\kappa_{\rm R}}{c}{\bf F}\cdot\mathbf{v}.
    \label{eq:total_energy} 
\end{aligned}
\end{equation}

\subsection{Opacities}
\label{sec:opacity}

The tabulated opacity is adopted from \citet{may05} where Planck and Rosseland
mean opacities for primordial matter including all three elements (H, He, and Li) are calculated. 
These opacities include the contribution from H species, namely, H, H$^-$, H$^+$, H$_2$, H$_2^+$, 
H$_3^+$, and D, He and Li species.   

The opacity tables cover the density range $-16 < {\rm log}\, \rho\, {\rm (g\,cm^{-3})} < -2$ and the 
temperature range $1.8 < {\rm log}\, T (K) < 4.5$. In our simulations, the gas collapse causes the density to
increase to $10^{-6} \,{\rm g\,cm^{-3}}$, and the temperature to increase above $2\times 10^4$\,K.  We have
extrapolated the temperature-depencence of the opacity by using the free-free, bound-free and electron scattering opacities. 

\subsection{Cooling and heating rates}
\label{sec:cool-heat}

For the optically-thin part of the collapse, we follow \citet{luo16}. The gas is assumed to
be dust free. In the optically-thick part of the flow, we have assumed LTE. 

To separate the optically-thin from thick regions, we have used the following complementary approaches. 
For adiabatic runs,
the optical depth $\tau$ is obtained using the Jeans length, $\lambda_{\rm J}$ for each cell, i.e.,
$\tau=\kappa\lambda_{\rm J}$, where $\kappa$ is the absorption opacity coefficient, calculated here
as the Planck mean, $\kappa_{\rm P}$. 

For the FLD runs, the position of the photosphere is calculated by tracing rays away from the 
densest cell
to a distance of 1\,pc, then integrating inwards to the point $\tau=1$, again using the
Planck mean opacity coefficient, $\kappa_{\rm P}$. We use 4,900 rays
equally spaced in azimuthal and polar directions. The resulting photosphere has no particular 
symmetry and its shape evolves each timestep. 

Furthermore, as a separate check to position the photosphere, we have used the values of the limiter, 
$\lambda=1/3$
(section\,\ref{sec:rad-transf}) as a trace of the optically-thick region. Both methods have been tested
and produced quite similar photospheric shapes and slightly different radii, with $\tau=1$
contour lying outside.

In order to compare  our FLD models with other models in the literature, we run models with an adiabatic 
equation of state. We have calculated the optical depth, $\tau$, over the Jeans length for each 
cell, and impose an exponential cutoff in the optically-thin cooling rate,
\begin{equation}
\label{eq:CoolCutoff}
  \Lambda = \Lambda_{\rm thin}{\rm e}^{-\tau},
\end{equation}
where $\Lambda_{\rm thin}$ is the optically-thin cooling rate.

\subsection{Initial Conditions}
\label{sec:ICs}

We have used initial conditions for isolated DM halos in this paper as described below. Fully cosmological 
initial conditions for our runs are presented in the companion paper. 
For the setup of isolated models, we follow the prescription developed by \citet{choi13}  
\citep[see also][]{luo16}.

We adopt the WMAP5 cosmological parameters \citep[][]{kom09}, namely, $\Omega_{\rm m} 
= 0.279$, $\Omega_{\rm b} = 0.0445$, $h = 0.701$, where $h$ is the Hubble constant in units of 
$100\,{\rm km\,s^{-1}\,Mpc^{-1}}$. We set up the details
of an isolated DM halo that is consistent with the cosmological context that we work with. Therefore, some
of the halo parameters are specified with units that include the Hubble parameter, although we use physical
quantities (not comoving) in this case. A DM halo is defined having
density equal to the critical density of the universe times the overdensity $\Delta_{\rm c}$,
which depends on
$z$ and the cosmological model. The top-hat model is used to calculate $\Delta_{\rm c}(z)$, 
and the density is calculated within a virial radius, $R_{\rm vir}$. The halo virial mass is 
$M_{\rm vir}(z) = (4\pi/3)\Delta_{\rm c}(z)\rho_{\rm c}R_{\rm vir}^3$. Because we treat the halos as
being isolated, all the values are calculated assuming $z=0$.

We  work  in  physical  coordinates and assume that the gas fraction in the model 
is equal to the universal ratio. The gas evolution is followed within DM halos of a virial mass of
$M_{\rm vir} = 2\times 10^8h^{-1}\,M_\odot$ and $R_{\rm vir} = 945h^{-1}$\,pc. The initial temperature of 
the gas is taken to be the virial temperature $T = 3.2\times 10^4$\,K. The simulation domain is a  box  
with a size $L_{\rm box} =  6$\,kpc centered  on  the  halo.   

The initial DM and gas density profiles are given by Eqs.\,1 and 2 of \citet{luo16}.  
The DM halo rotation is defined in terms of the cosmological spin parameter $\lambda=J/\sqrt{2}M_{\rm vir}
R_{\rm vir} v_{\rm c}$, where $J$ is the angular momentum of the DM halo, and $v_{\rm c}$ is the circular 
velocity at $R_{\rm vir}$ \citep[e.g.,][]{bull01}. We use $\lambda=0.03$.
 
To produce DM halos with a pre-specified $\lambda$ for isolated  halo  models, we follow the 
prescription  of  \citet{long14} and \citet{col18}. In short, we assume a
DM velocity distribution with an isotropic velocity dispersion, then reverse the tangential velocities of
a fraction of DM particles (in cylindrical shells) to 
obtain $\lambda$ equal to the required value. This action
preserves the solution of the Boltzmann equation and is a direct corollary of the \citet{jeans19}
theorem \citep[e.g.,][]{lynd60,binn08}.

For the gas in AMR grid  cells,  we  calculate  the  average  tangential  velocities  of
the background DM in cylindrical shells, accounting for the
dependence along the (rotation) $z$-axis. The radial profile of the DM tangential velocity is
given by Eq.\,4 of \citet{luo16}.  

The  DM spatial resolution  is adaptive and set  by  the gravitational  softening  length, corresponding to the cell size. For the initial root grid of $64^3$ in a 6\,kpc region with 
a maximal refinement level of 8 allowed for gravity from the DM particles, $\epsilon_{\rm DM,min} = 
6000/64/2^8 = 0.37$\,pc. This value is kept constant.

For the gas, the gravitational softening is adaptive with the maximal refinement level of 33. However,
in all simulations, only a refinement level of 30 has been reached.
We use the initial resolution of $100^3$ particles-in-mesh for the DM. The force resolution in adaptive  
PM codes  is  twice  the  minimal cell size \citep[e.g.,][]{kra97}.

\section{Results: Adiabatic flow}
\label{sec:results_adia}

\begin{figure}
\center
\includegraphics[width=0.5\textwidth,angle=0] {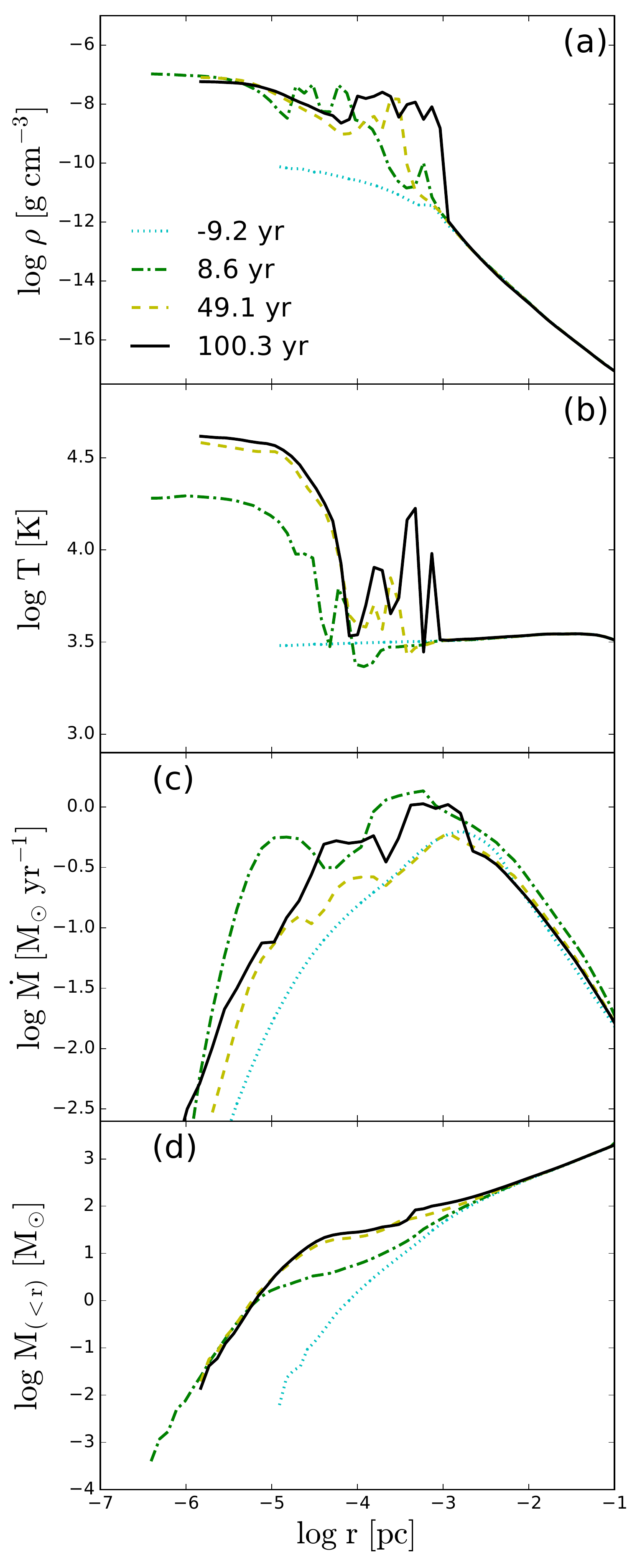}
\caption{Adiabatic accretion flow. Evolution of (a) gas density, (b) temperature, (c) accretion rate,
and (d) mass within spherical radius $R$, at a few representative times. Negative values correspond to
times prior to the establishment of the photosphere.
}
\label{fig:adia_profile1}
\end{figure}

We start by presenting results of adiabatic runs of direct collapse within isolated halos. The FLD models
are presented in the next section. The early stages of the gravitational collapse have
been simulated here but discussed elsewhere \citep{choi13,choi15,shlo16}. Here, we redefine the $t=0$ time at a much later stage
and focus on the innermost regions, $\ltorder 0.1$\,pc, of the collapse. This happens at $\sim 1.993$\,Myr
after the start of simulation. Times prior to this point are specified as negative.  
We find it convenient to choose this time when the flow forms a `photosphere' 
(section\,\ref{sec:cool-heat}),
which corresponds to the time when the optical depth in the flow becomes larger than unity. This definition
differs from the density cutoff which is used in some publications.

The optically-thin part of the collapse exhibits a self-similar, Penston-Larson profile of $\rho\sim
R^{-2}$ \citep{luo16}. The small degree of a rotational support in the halo does not modify this behavior for the first 1--1.5
decades in radius. For the isolated models presented here, the angular momentum is nearly conserved
in the outer region, due to the idealized initial conditions. This leads to a slowdown in the collapse
inside $\sim 10$\,pc due to the angular momentum barrier, which can be observed in the density, temperature 
and velocity distributions, in agreement with \citet{choi13}. A standing shock forms and
leads to a substantial decrease in the mass accretion rate there and to a mass accumulation.
With the exception of this shock, the gas stays nearly isothermal, with $T\sim 3,000-5,000$\,K.

On spatial scales $\sim 0.3-3$\,pc, the density ratio of $\rho_{\rm gas}/\rho_{\rm DM}$ increases
and reaches unity. The gas effectively decouples from the DM interior to these radii. Still, the
DM can exert gravitational torques on the interior gas and absorb its angular momentum.

On scales of $\ltorder 10$\,pc, within the disk-like configuration, the collapse is dominated by the 
Fourier mode $m=2$, and the accretion flow exhibits a density enhancement in the form of a filament which
can be traced as deep as  $\sim 10^{-5}$\,pc. This shows that
even the innermost flow remembers the physical conditions at larger scales.
The evolution of the basic parameters of the accretion flow inside 0.1\,pc is shown in 
Figure\,\ref{fig:adia_profile1}.

\begin{figure}
\center
 \includegraphics[width=0.5\textwidth,angle=0] {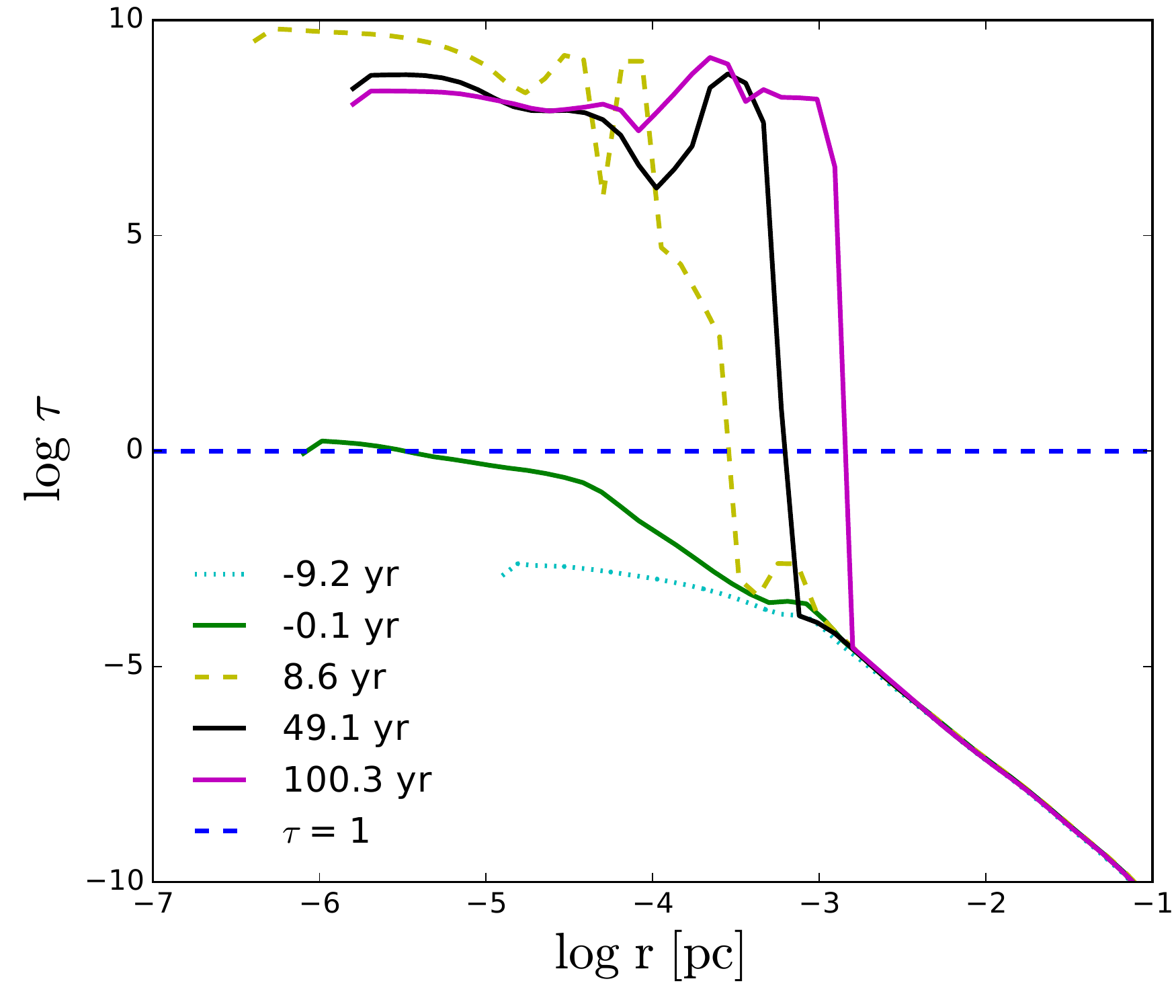}
\caption{Adiabatic collapse: optical depth profile in the flow at a few representative times. Negative values
correspond to times prior to the establishment of the photosphere. The dashed horizontal line delineates the photosphere 
at $\tau=1$.
}
\label{fig:tau}
\end{figure}

\begin{figure}
\center
\includegraphics[width=0.45\textwidth,angle=0] {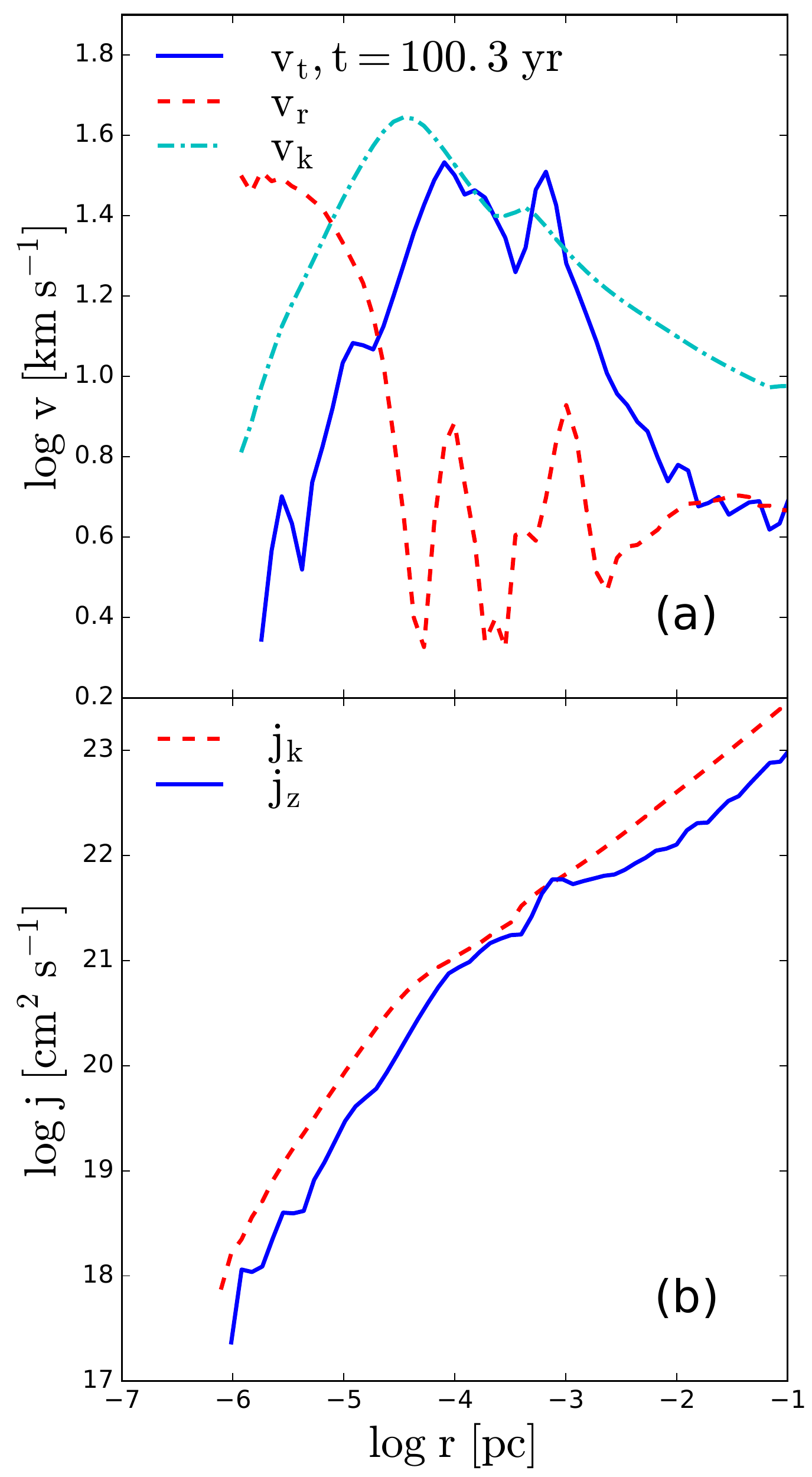}
\caption{Isolated adiabatic accretion final profiles at $t=100.3$\,yr:
(a) tangential velocity, $v_{\rm t}$ (solid line), radial inflow velocity, $v_{\rm r}$ (dashed line), and
circular velocity, $v_{\rm k}$ (dot-dashed line) at cylindrical radius $r$;
(b) specific angular momentum of accreting gas, $j_{\rm z}$ (solid line) and circular specific angular momentum, 
$j_{\rm k}$ (dashed line) at $r$. 
}
\label{fig:adia_profile2}
\end{figure}

We have introduced a cutoff in the cooling rate of the flow based on its optical depth
(Eq.\,\ref{eq:CoolCutoff}). Above $\tau=1$, the cooling rate decreases exponentially. This cutoff
mimics the formation of the photosphere, below which the radiation is expected to diffuse rather than 
free-stream, 
and the cooling rate is expected to decrease sharply. Very roughly, this condition is fulfilled initially at
$R_{\rm ph}\sim 10^{-6}$\,pc, and this radius expands rapidly to $R_{\rm ph}\sim 10^{-4}-10^{-3}$\,pc 
(Fig.\,\ref{fig:tau}). The FLD model, described in the next section, behaves similarly.  

The flow quickly becomes adiabatic interior to this radius, which can be observed by monitoring the
cooling rate. Outside $R_{\rm ph}$, we observe the radiative cooling, $\Lambda$, being compensated 
by compressional heating. Inside  $R_{\rm ph}$, on the other hand, the
compressional heating dominates, resulting in a steep rise in temperature. 

Figure\,\ref{fig:adia_profile1}a shows the density profile at four representative times during the collapse.
The region where the flow becomes optically thick, $R_{\rm ph}$, displays a sharp increase in the gas density, 
which levels off at smaller radii. Note the formation of a central core with $R\sim 10^{-5}$ pc at 
$t\sim 8.6$\,yr, and a number of density peaks outside the core at later times, which represent the 
forming fragments, as we discuss below. The temperature profile is closely related to the formation of the
core and surrounding fragments (Fig.\,\ref{fig:adia_profile1}b). The central density and temperature have
reached $\sim 10^{-7}\,{\rm g\,cm^{-3}}$ and $5\times 10^4$\,K, respectively. The fragments stand out clearly 
at the end of the simulation 
as temperature peaks. By the end, both core radius and the new fragments are situated at larger radii, 
as $R_{\rm ph}$ has moved out.

\begin{figure*}
\centerline{
\includegraphics[width=1.0\textwidth] {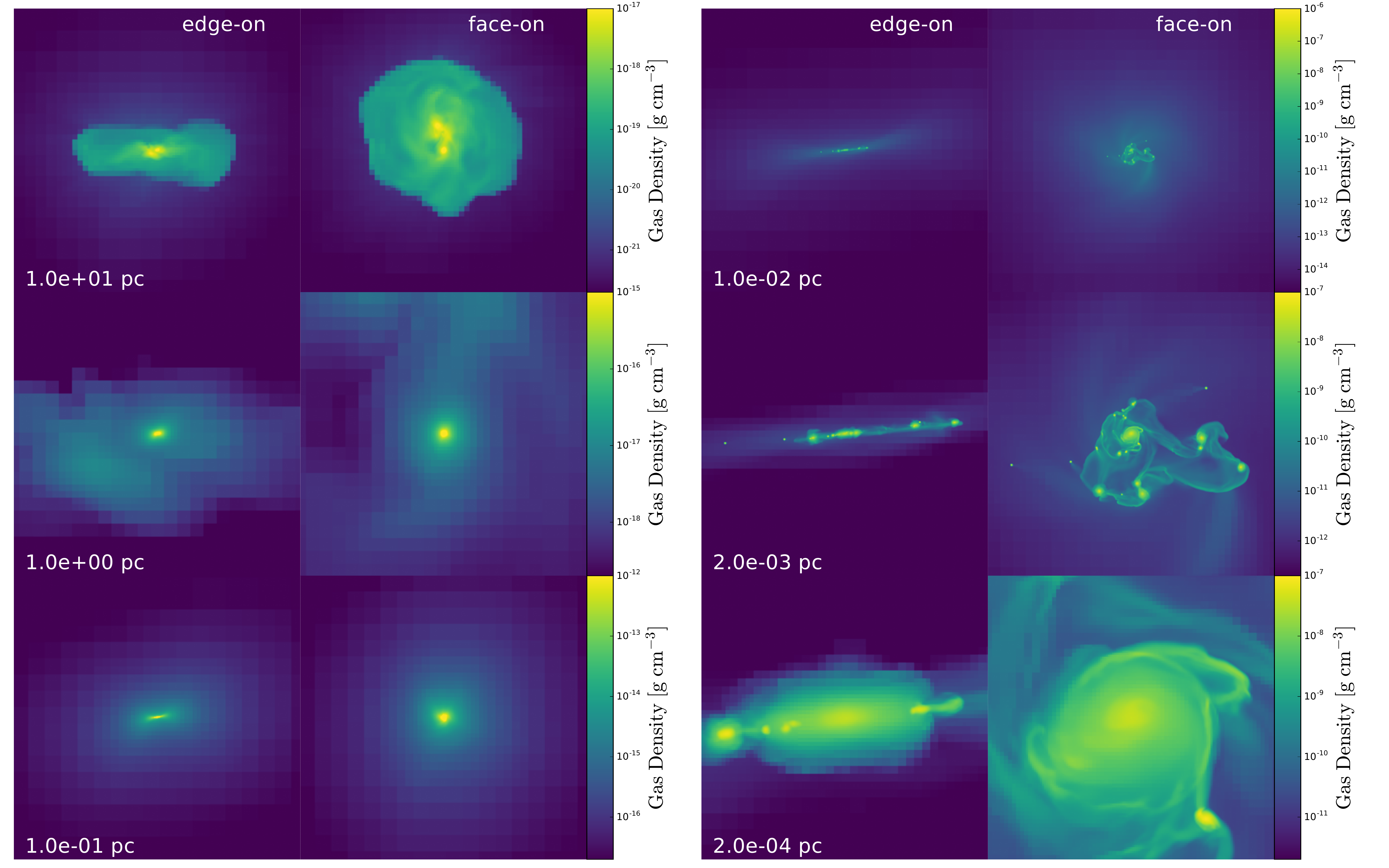}
}
\caption{Final projection snapshots of adiabatic collapse on various spatial scales, from
10\,pc down to $2\times 10^{-4}$\,pc. Shown are two independent directions, roughly corresponding
to face-on and edge-on views. Fragmentation is occurring on scales of $\sim 10^{-3}$\,pc, somewhat
larger than the photospheric scale. 
}
\label{fig:adia_final_proj}
\end{figure*}

The mass accretion rate profile reflects the existence of a disk on scales of $1-10$\,pc, shown in
Figure\,\ref{fig:adia_final_proj}, which is largely rotationally supported. The inner parts of
this disk become unstable and collapse, with accretion rate ${\dot M}(R)$, reaching a maximum
and declining further inwards (Fig.\,\ref{fig:adia_profile1}c).
The shape of ${\dot M}(R)$ stays largely unchanged with time, except for some variation
of the peak position, which shifts back and forth. 

\begin{figure}
\center
 \includegraphics[width=0.47\textwidth,angle=0] {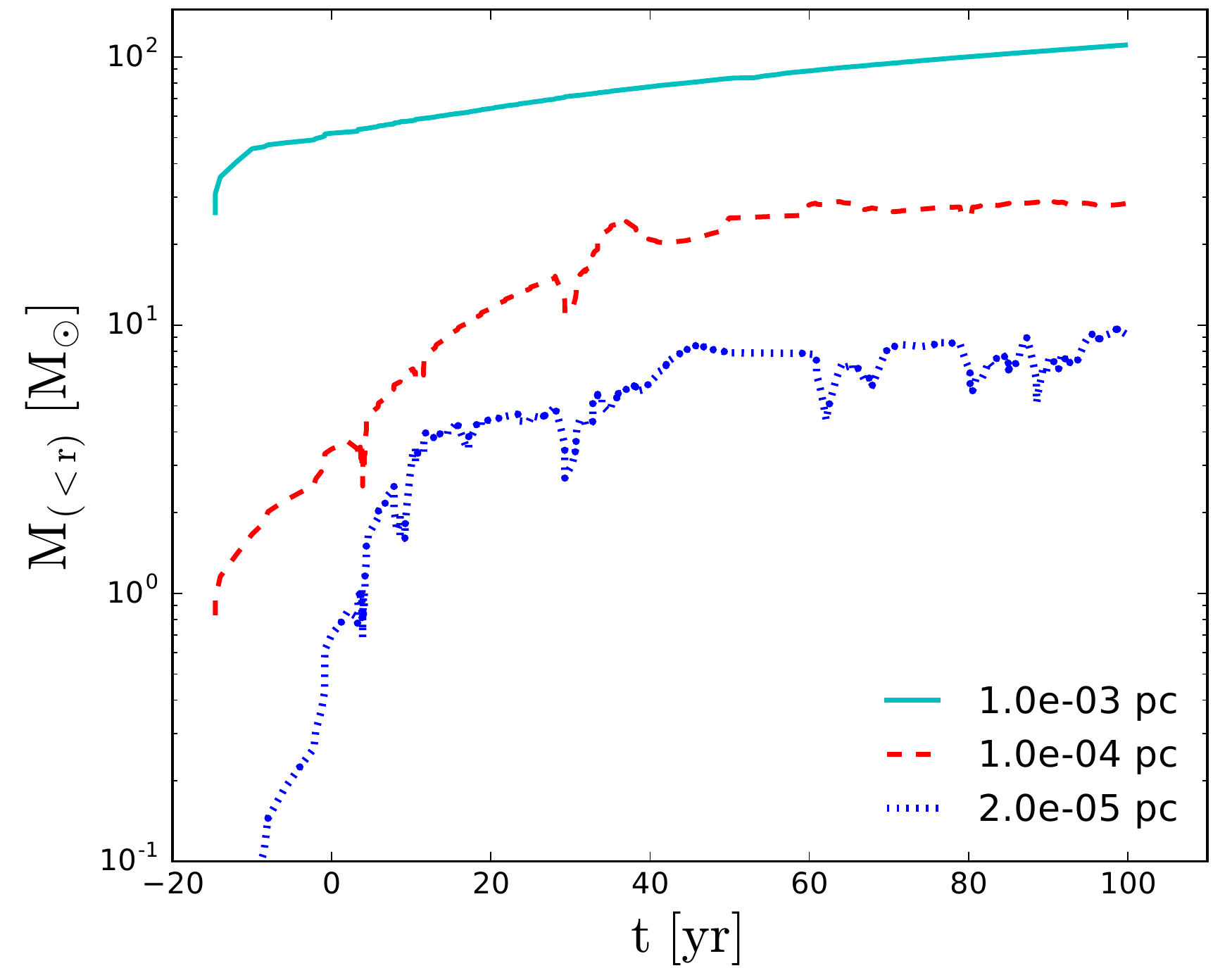}
\caption{Adiabatic collapse: evolution of the enclosed mass within fixed spherical radii.
}
\label{fig:encl_mass}
\end{figure}

As expected and despite formation of a disk at larger radii, the mass accumulates within the central region
(Fig.\,\ref{fig:adia_profile1}d). By $t\sim 100$\,yr, the amount of gas within the central $\sim 10^{-4}$\,pc
is $\sim 40\,M_\odot$, and within 0.1\,pc about $2\times 10^3\,M_\odot$. The mass within the photospheric
radius is $\sim 100\,M_\odot$ (Fig.\,\ref{fig:encl_mass}).

The rotational support on small scales, within $R_{\rm ph}$, is partial but prominent in the adiabatic flow
(Fig.\,\ref{fig:adia_profile2}b). The flow is rotationally supported, within a factor of 2, at nearly all
radii, but gains even more support around $\sim 10$\,pc, where it forms a warped disk as discussed above, and 
around $R_{\rm ph}$, where it forms a growing geometrically-thick disk surrounded by fragments at the later 
stage (e.g., Fig.\,\ref{fig:adia_final_proj}).  The fragments can also be traced in the $\rho$ and $T$ 
distributions averaged on spherical shells in Figure\,\ref{fig:adia_profile1}a,b. By $t\sim 100$\,yr, tangential 
velocity rises to its maximal value at $\sim R_{\rm ph}$, then decreases by about a factor of 10, and 
the radial velocity behaves in the opposite way (Fig.\,\ref{fig:adia_profile2}a). What is the reason for 
this decrease in $v_{\rm t}$ and increase in $v_{\rm r}$ at smaller $R$? We analyze this issue below.

\begin{figure*}
\center
\includegraphics[width=1.0\textwidth,angle=0] {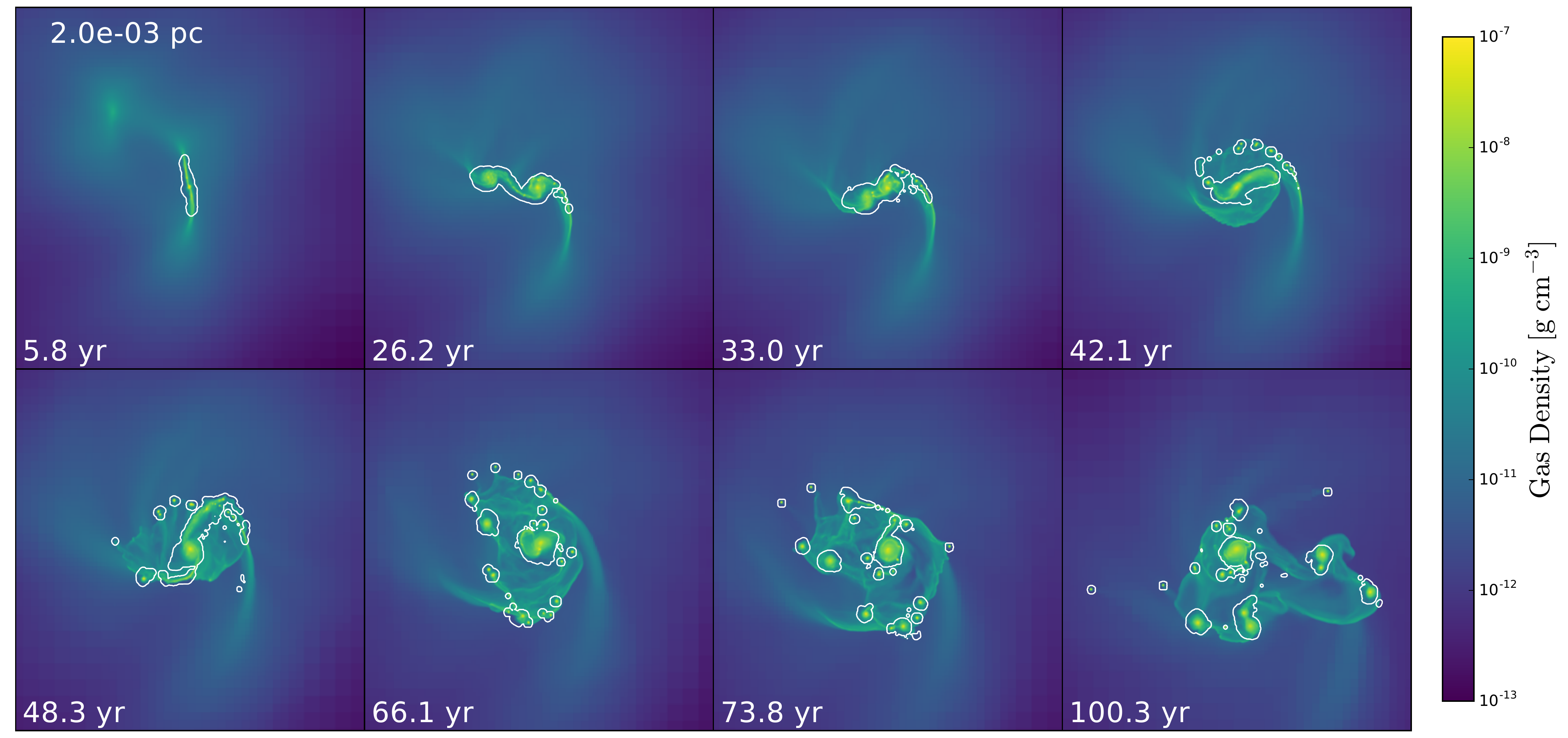}
\caption{Evolution of adiabatic collapse. Projection snapshots on scale of $2\times 10^{-3}$\,pc. Color 
palette is based on logarithmic scale. The white contours represent the photospheric surfaces defined in 
the text.
}
\label{fig:adia_proj_evol}
\end{figure*}

Evolution of the accretion flow on scales of $\sim 10^{-3}$\,pc reveals a dominant bar-like mode at early
times (e.g., Fig.\,\ref{fig:adia_proj_evol}). At a later stage, $t\gtorder 26$\,yr, two open spiral arms are 
driven by this bar-like feature and completely dominate the flow. Fragmentation is seen in projection. Most 
of the fragments spiral in and merge in the central region.

Figure\,\ref{fig:adia_final_proj} provides more details about the central region of $\sim 10^{-4}$\,pc,
where one observes a disk, edge-on and face-on, at the end of the run. 
Fourier analysis of this disk reveals a strong bar-like mode dominating its kinematics, with an amplitude of 
$A_2\sim 0.46$. The $m=1$ mode is less
important. We have also measured the strength of the gaseous bar using its ellipticity, defined as 
$\epsilon=1-b/a$, where $a$ and $b$ are the semi-major and semi-minor axes \citep[e.g.,][]{marti06}. The 
typical value for the late stage is $\epsilon\sim 0.65$, which means that a strong bar dominates the 
potential in this region. This $m=2$ mode 
leads to strong radial flows which explains the radial profiles of tangential and radial velocities.

\begin{figure}
\centerline{
 \includegraphics[width=0.5\textwidth,angle=0] {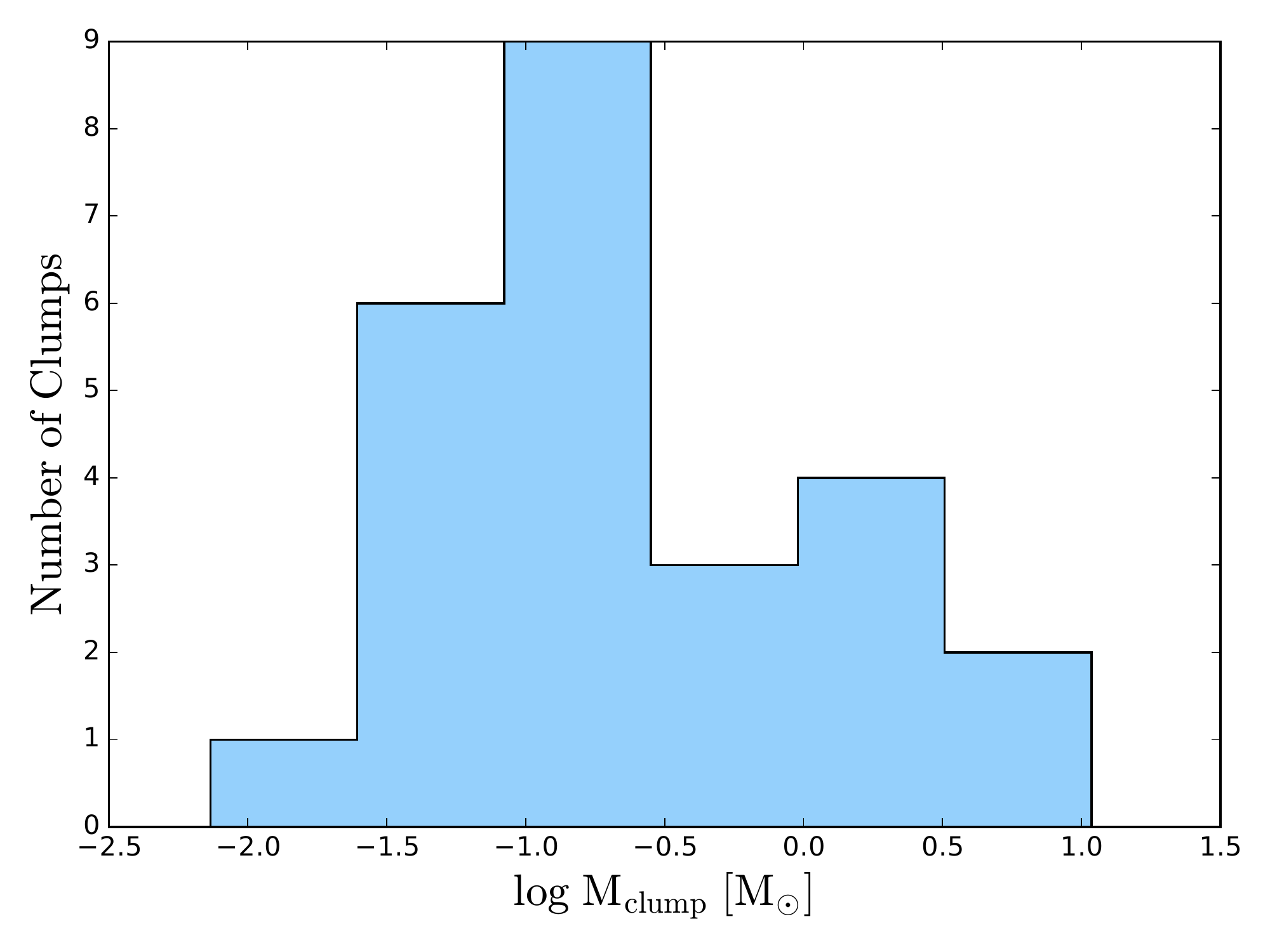}
}
\caption{Adiabatic collapse: fragmentation of the flow on scales of $\sim 10^{-3}$\,pc (see 
Fig.\,\ref{fig:adia_proj_evol}). The distribution of clump masses is shown.
}
\label{fig:clumps}
\end{figure}

The distribution of fragment masses is given in Figure\,\ref{fig:clumps}. These clumps have been identified
by having detached photospheres from the central object, then verified being self-gravitating. The most 
massive clump corresponds to 
the central disk, $\sim 10\,M_\odot$, and the majority of clumps have masses of $\sim 0.1\,M_\odot$. Their 
formation is limited to the region dominated
by the spiral arms, i.e., within $\sim 10^{-4}-10^{-3}$\,pc. In fact, these clumps form
along the spiral arms only. The clumps that formed earlier spiral in and are absorbed by the central disk. 
The number of clumps levels off in time, reaching a steady state.

To understand the reason for fragmentation, we have checked for Toomre instability, characterized
by the $Q=\chi\,c_{\rm s}/\pi G \Sigma$ parameter. Here $\chi$ is the epicyclic frequency, $c_{\rm s}$
is the sound speed, and $\Sigma$ is the surface density of the disky entity. Based on the properties
of the flow, we have calculated $Q(r)$ for $t=27$\,yr. We find that it dips below unity between
$10^{-4}$\,pc and $10^{-3}$\,pc from the center, i.e., exactly where the clumps are observed to form
(e.g., Fig.\,\ref{fig:adia_proj_evol}).

However, caution should be exercised here, as the clumps form in the spiral arms, while the underlying
disk is ill-defined. An alternative explanation can be related to the Kelvin-Helmholtz (K-H) shear
instability \citep[e.g.,][]{chan61}. The open spirals represent shock fronts, and the gas moves
through them with a Mach number of ${\mathcal M}\sim {\rm few}$, as can be inferred from 
Figures\,\ref{fig:adia_profile1} and \ref{fig:adia_profile2}. The gas experiences
an oblique shock, and the measured pitch angle between the shock front and the gas streamlines
is about $i\sim 60^\circ$, confirming that the spirals are open and not tightly wound.

Such a configuration will induce shear in the flow, close to the shock front, and may be subject to
Kelvin-Helmholtz shear instability, when the associated Richardson number, $Ri < 0.25$ \citep{chan61}. 
This instability will affect
the shock front which will `wiggle', and clumps will form and grow at the vertices of the distorted
shock front \citep[e.g.,][]{bal88,kim02}. The Richardson number is given by \citep{chan61},
\begin{equation}
Ri =  -\frac{g}{\rho} \frac{{\rm d}\rho/{\rm d}z}{\left({\rm d}v/{\rm d}z\right)^2},
\end{equation}
where the $z$ axis is directed perpendicular to the shock front, $g$ is the (self-) gravitational 
acceleration due to the shocked material, and $v$ is the shear velocity.
The gas self-gravity will act as a stabilizer, and its effect on the flow must be estimated.

We assume that the shock front and the postshock layer are associated with the spiral arm which 
perturbs an otherwise axisymmetric background gravitational potential. It is convenient to
estimate the value of the gravitational acceleration induced by the spiral arm as a fraction
$\beta$ of the radial potential measured by the centrifugal acceleration, $v_{\rm t}^2/R$
where $v_{\rm t}$ is the gas tangential velocity. $\beta\sim 0.05$ is a typical perturbation by a 
spiral arm in disk galaxies \citep[e.g.,][]{engl00}.
To project this acceleration on the direction of the streamlines entering the shock front, we account 
for the pitch angle $i$, to obtain $g\sim \beta v_{\rm t}^2/r {\rm sin}\,i$. Here $r$ corresponds
to the radius vector extending from the flow center, i.e., in our case, corresponding to the 
core center.

The shear velocity can be estimated from the velocity difference across the thickness of the shocked 
layer, $b$, projected onto the normal to the shock front, $v\sim v_{\rm t}{\rm cos}\,i$. Assuming that 
the pre-shock gas has a large Mach number, the Richardson number is
\begin{equation}
Ri\sim \beta \left(\frac{b}{r}\right) \left({\rm sin}\,i\,{\rm cos}^2\,i\right)^{-1}.
\end{equation}
 
Adopting values from the run, i.e., $i\sim 60^\circ$, and $b/r\sim 0.3$,  
we obtain $Ri\sim 0.1$. Hence the flow must be unstable and form clumps along the spiral arms.

Next, we invert the problem, and ask what wavelengths, $l/r$, are unstable, taking $Ri=0.25$ 
and keeping other values fixed. For K-H shear instability, we obtain $l/r < 0.8$.  

Hence, the K-H shear instability appears as an viable alternative to the Toomre's instability, especially 
because the fragmentation happens in the spiral arms and the underlying disk is ill-defined. The latter
comment refers to the formation of spirals in a sheared flow dominated by a bar-like feature.

The central object, which is supported mainly by rotation and partly by gas pressure gradients, does not
show any tendency to fragmentation. This is understandable, because it is geometrically thick and 
Toomre instability is suppressed with increasing thickness, in contrast to the claim by \citet{bec15}. 

To demonstrate the mass growth in the central region, we measure the mass evolution contained within 
three specific radii, i.e., $10^{-5}$\,pc, $10^{-4}$\,pc, and $10^{-3}$\,pc (Fig.\,\ref{fig:encl_mass}). 
For the largest radius, $10^{-3}$\,pc, the growth is monotonic,
and the accumulated mass is about $100\,M_\odot$.
The noise increases gradually to smaller radii. The amount of gas within $10^{-4}$\,pc radius appears 
to saturate in time, which
is explained by the fragmentation. The fragments spiral in more slowly than the smooth accretion flow, and are 
responsible for the mass accumulation on this scale. 

To summarize, we clearly observe formation of a central object in the adiabatic accretion flow. 
This object appears to be supported by both the gas thermal pressure and rotation. The radiation pressure
gradients are not important, as the temperature remains relatively low, $T < 10^5$\,K.
Core formation results in a substantially flattened configuration, resembling a geometrically-thick
disk. It is surrounded by fragments and we lean towards the K-H shear instability explanation for
their origin, as opposed to Toomre instability. We return to this issue in the Discussion section.

\section{Results: non-adiabatic flow}
\label{sec:results}

One of the main questions about direct collapse is whether the adiabatic flow approximation used in the 
literature so far adequately represents evolution. In this sense, the adiabatic run with identical initial 
conditions serves as a test model of what to expect in the FLD run.
The non-adiabatic flow in the isolated model is in many respects similar to that of the adiabatic model, 
but also exhibits some important differences that cannot be ignored. As noted earlier, we consider the FLD 
flow to be in LTE. 

\begin{figure}
\center
\includegraphics[width=0.43\textwidth,angle=0] {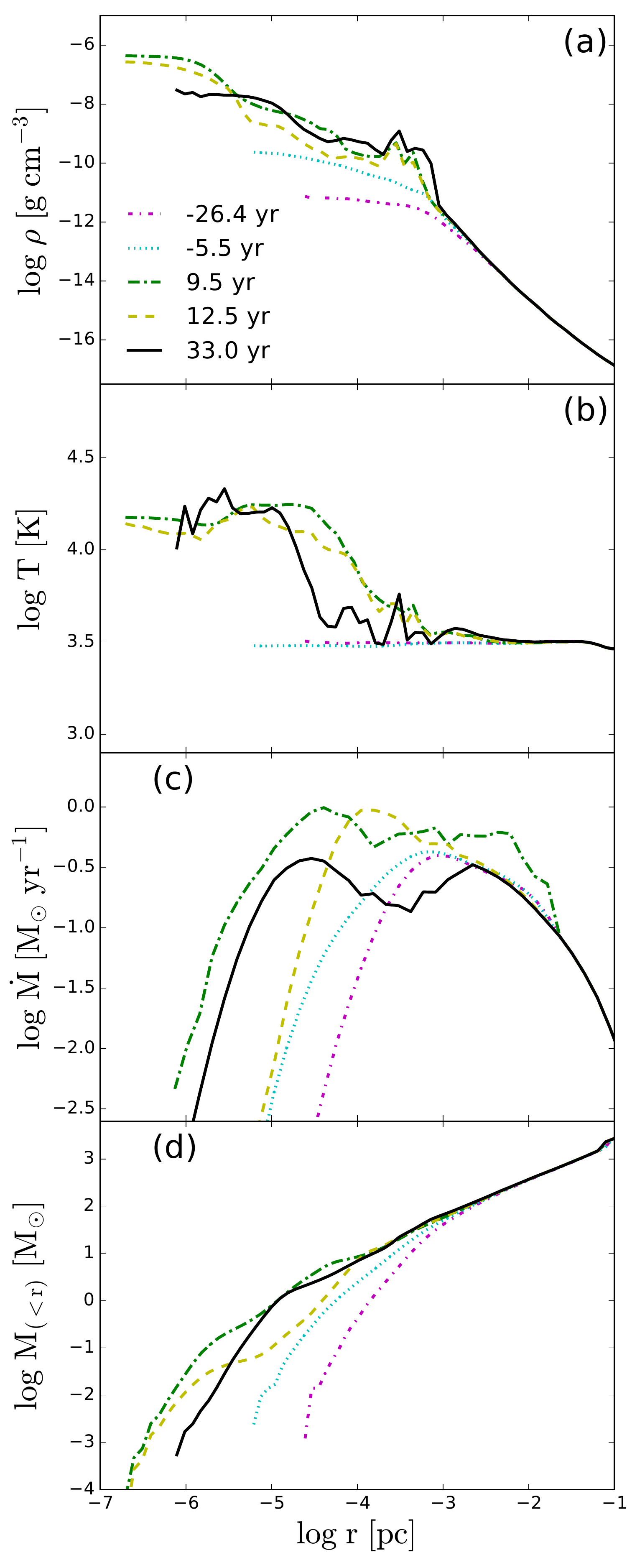}
\caption{Non-adiabatic accretion flow. Evolution of (from top to bottom): gas density, temperature, accretion 
rate, and mass within spherical radius $R$.
}
\label{fig:FLD_profile1}
\end{figure}

\subsection{Deep interior flow: Formation and dissolution of the central core}
\label{sec:interior}

\begin{figure*}
\centerline{
 \includegraphics[width=1.0\textwidth,angle=0] {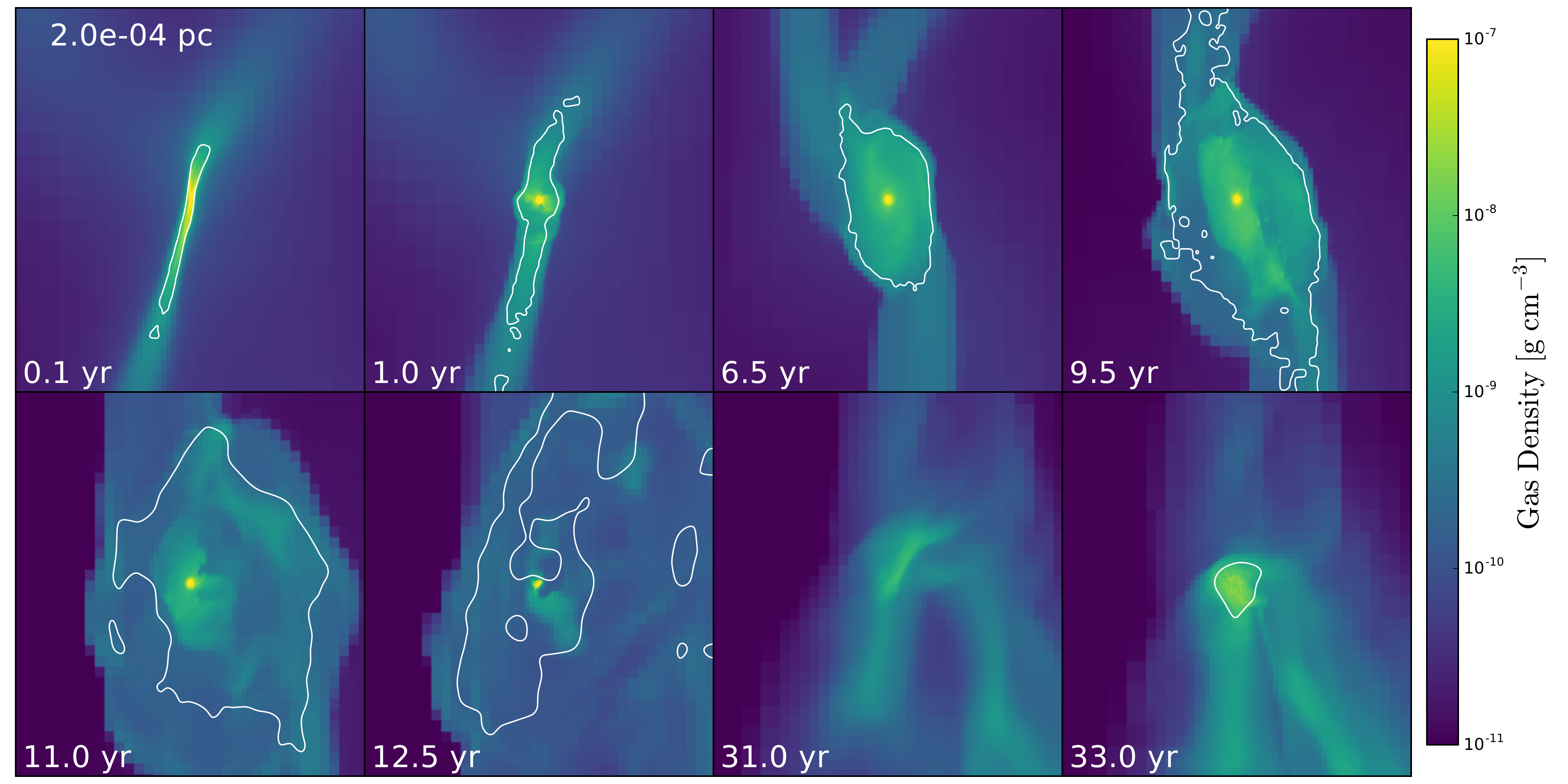}
}
\caption{Evolution of non-adiabatic collapse. Projection snapshots on scale of $2\times 10^{-4}$\,pc. Color 
palette is based on logarithmic scale. The contour line corresponds to the position of the photospheric 
surface and was calculated using the delimiter $\lambda=1/3$ (section\,\ref{sec:rad-transf}). Note that 
after $t\sim 9.5$\,yr, the photosphere is expanding because of the extensive outflow from the central core 
region, then receding. The core dissolves completely by $t\sim 15$\,yr, and the region becomes marginally 
optically-thin.
A new core starts to form at around this time in a slightly different position and reaches 
$\sim 1\,M_\odot$ by $t\sim 32$\,yr. Frames before $t=15$\,yr have been centered on the existing core,
while later frames are centered on the forming new core (see also Figures\,\ref{fig:FLDproj2}  and 
\ref{fig:FLD_Lrad}).
}
\label{fig:FLD_Rph}
\end{figure*}

The basic parameters of the FLD flow are shown in Figure\,\ref{fig:FLD_profile1}. They display the
gradual formation of the central object, its photosphere, and its subsequent expansion. The photospheric
jump is not as dramatic as in the adiabatic case. The central density is higher at some specific time
in the run, then becomes lower. The photosphere forms at 2.037\,Myr after the start of the simulation,
and $t=0$ occurs slightly later in the evolution compared to the adiabatic run, by about $10^4$\,yr.
The central temperature is lower than in the adiabatic case and remains stable after 
the formation of the photosphere, $\sim 1.6\times 10^4$\,K. Fragmentation is virtually non-existent,
and formation of a few photospheric `islands' is observed, that merge quickly.
The FLD flow is filamentary, as in the adiabatic case, with a single dominating filament, as seen in 
Figure\,\ref{fig:FLD_Rph} at early times.

The temperature starts to rise at $\sim {\rm few} \times 10^{-4}$\,pc (Fig.\,\ref{fig:FLD_profile1}b). The
rise from $T\ltorder 3\times 10^3$\,K correlates with a sharp increase in the bound-free
opacity in atomic hydrogen. Photoionization quickly becomes the dominant heating mechanism in the gas,
surpassing the compressional heating by orders of magnitude. This leads to a sharp increase in the 
optical depth and to the appearance of the photosphere at $\tau\sim 1$ which we denote as $R_{\rm ph}$
as in the adiabatic case. This time is taken as $t=0$.

\begin{figure}
\center
\includegraphics[width=0.43\textwidth,angle=0] {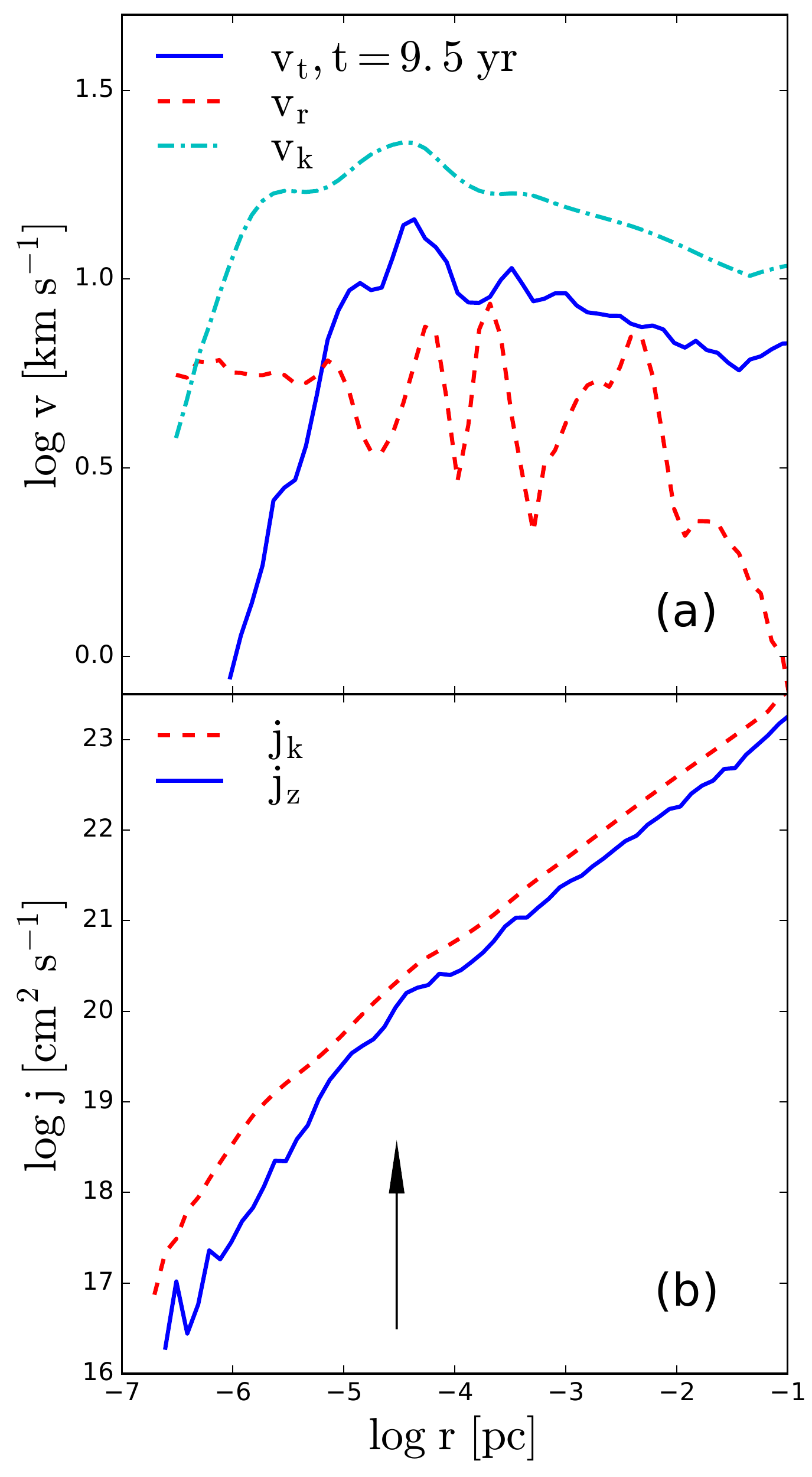}
\caption{Non-adiabatic accretion flow final profiles of
(a) tangential velocity, $v_{\rm t}$ (solid line), radial inflow velocity, $v_{\rm r}$ (dashed line),
and circular velocity,  $v_{\rm k}$ (dot-dashed line) at cylindrical radius $r$;
(b) specific angular momentum of accreting gas, $j_{\rm z}$ (solid line) and circular specific angular 
momentum, $j_{\rm k}$ (dashed line) at $r$. The vertical arrow shows the approximate position of the
photosphere.
}
\label{fig:FLD_profile2}
\end{figure}

\begin{figure}
\centerline{
\includegraphics[width=0.48\textwidth,angle=0] {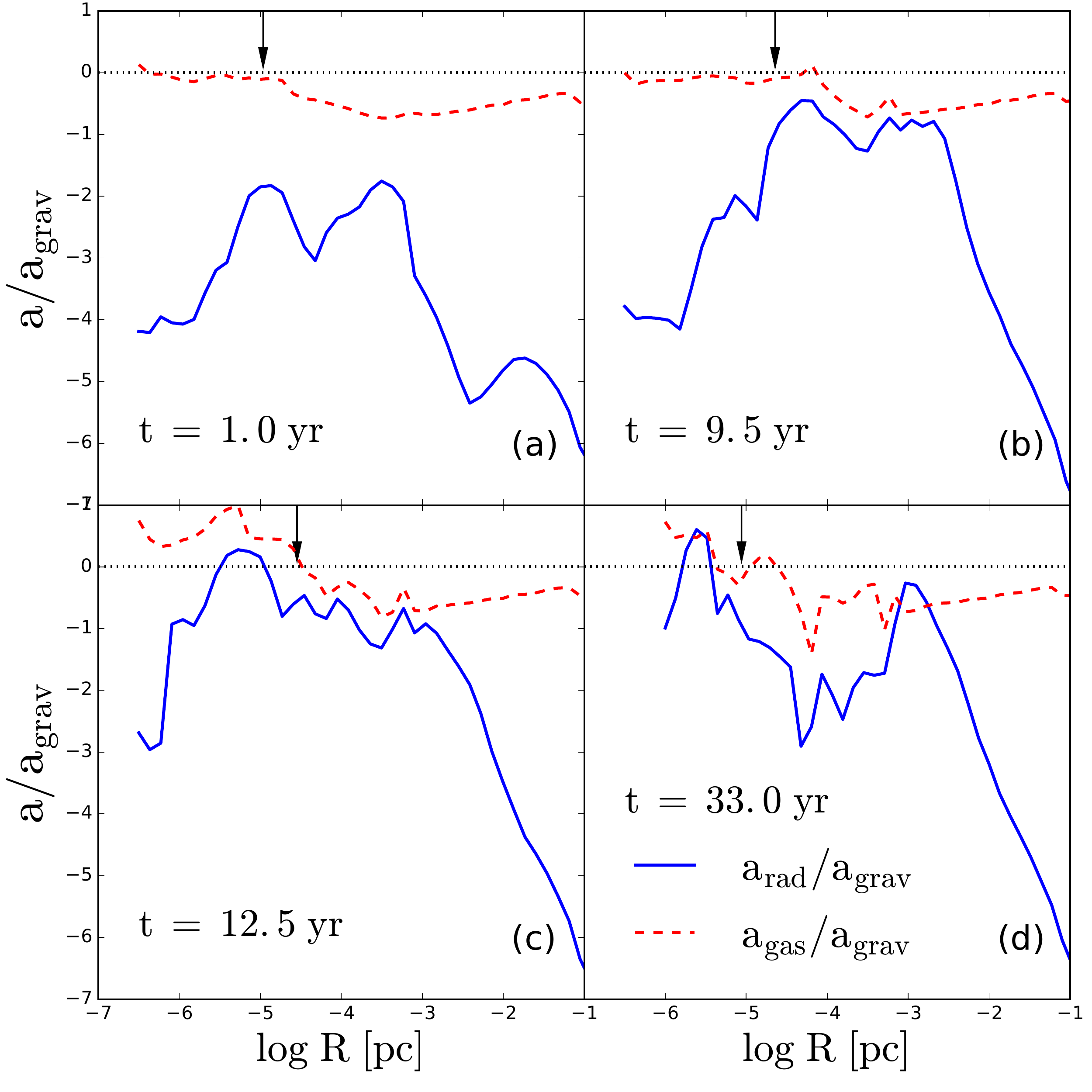}
}
\caption{Dominant accelerations in the non-adiabatic accretion flow: thermal pressure gradient (dashed red 
line) and radiation pressure gradient (solid blue line) normalized by gravitational acceleration of the 
enclosed mass at $t\sim 1$\,yr, 6.5\,yr, 12.5\,yr and 32.7\,yr. The dotted line is drawn to delineate the ratio
of unity. The vertical arrows show the approximate position of the photosphere.
}
\label{fig:FLD_accel}
\end{figure}

The density profile within $R_{\rm ph}$
becomes flatter than $R^{-2}$, and is rather closer to $R^{-1}$
(Fig.\,\ref{fig:FLD_profile1}a). Initially, the collapse proceeds deeper than in the adiabatic case, down
to $\sim 10^{-7}$\,pc, before it is stopped by the gas pressure gradient. The radiation force is
about 1\% of the gas pressure force at this time (Fig.\,\ref{fig:FLD_accel}a), then increases to about 
10\% by $t\sim 9.5$\,yr, and continues to increase thereafter (Fig.\,\ref{fig:FLD_accel}b).
Rotation is partially important at $R_{\rm ph}$, but declines sharply at smaller radii 
(Fig.\,\ref{fig:FLD_profile2}a,b).

The evolution starts to diverge from the adiabatic flow at small radii. Within $R_{\rm ph}$, a core
forms and grows to $\sim 1\,M_\odot$ and size $\sim 7-8\times 10^{-6}$\,pc. Its temperature is lower
than the outside gas by a factor of 2, and its density increases.
One can observe the associated break in the density profile of Figure\,\ref{fig:FLD_profile1}a. 

The filamentary inflow develops as the collapse proceeds and extends to the smallest scales achieved in 
the run. One observes that the inflow is channeled along these filaments, and outside material joins 
the filaments
after experiencing an oblique shock on their surfaces. Additional shocks form in the central region
where the two main filaments collide and the flows merge. Velocities abruptly decrease within the 
innermost shock,
pointing to the overall slowdown of the accretion flow, and the start of virialization. Both the thermal 
pressure gradients in the gas and the rotational support contribute to this dramatic slowdown and
essentially terminate the accretion flow.

\begin{figure*}
\centerline{
 \includegraphics[width=1.0\textwidth,angle=0] {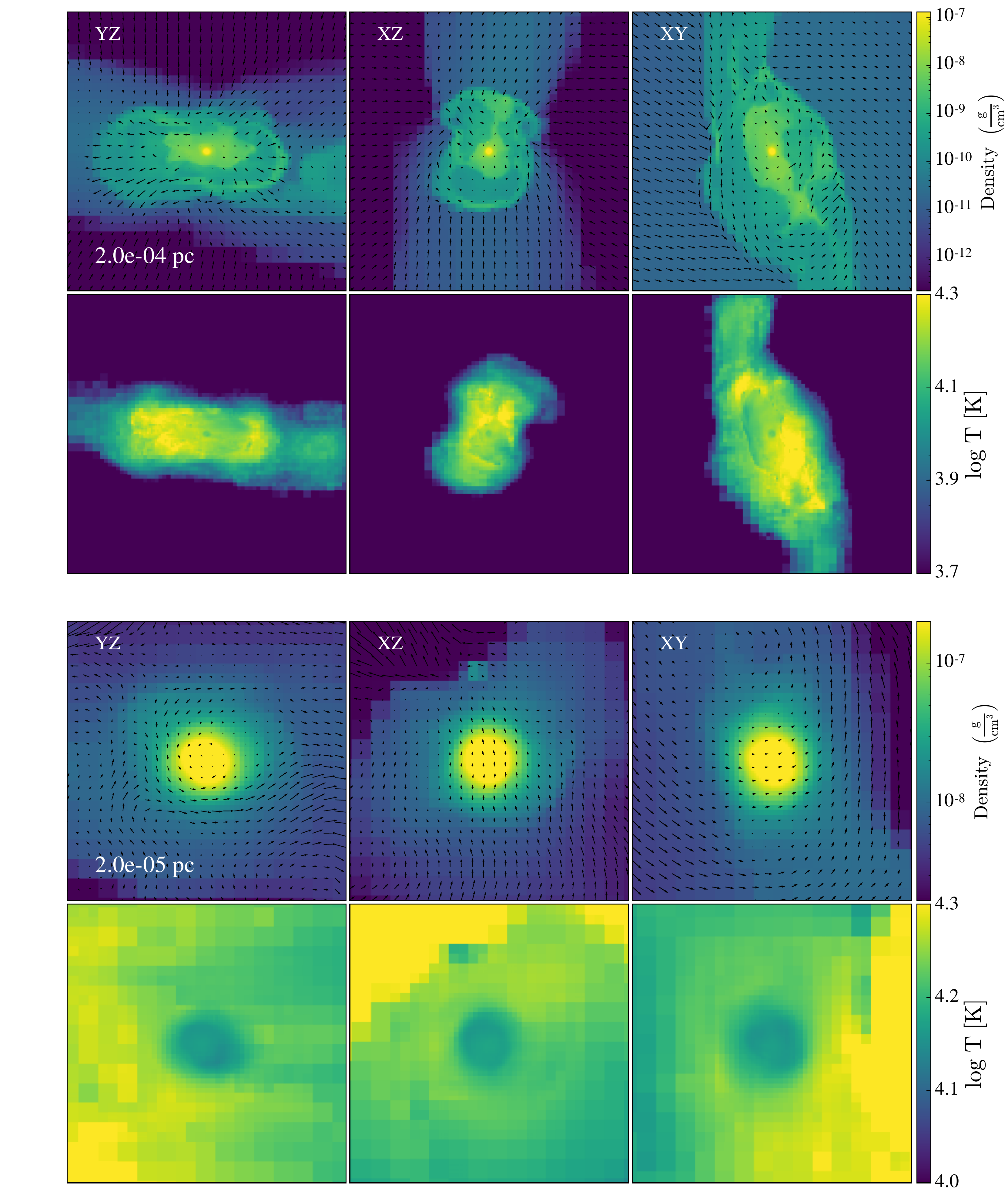}
}
\caption{Non-adiabatic accretion final projection snapshots from three independent directions on scales 
of $2\times 10^{-4}$\,pc (top) and  $10^{-5}$\,pc (bottom) at $t\sim 9.5$\,yr (see corresponding
Figure\,\ref{fig:FLD_Lrad}a frame at this time).
On each scale we show the density and velocity fields projections (top) and the temperature (bottom).
Note the developing anisotropic outflow from the central core along the filament and the associated 
expanding bubble driven by this outflow. The shape of the core is clearly outlined by the large density
contrast with the environment. Its interior temperature is slightly lower than that of the surrounding 
gas.
}
\label{fig:FLDproj2}
\end{figure*}

To understand the prevailing structure on these scales, refer to Figure\,\ref{fig:FLDproj2}, 
which provides
views of the region on scales $R\sim 2\times 10^{-4}$\,pc (top frames) and $R\sim 2\times 10^{-5}$\,pc 
(bottom frames), at $t\sim 9.5$\,yr. Density, temperature and flow pattern are shown. On smaller scales, 
one observes a small dense core which is nearly round, confirming the unimportance of rotation. This core is
surrounded by a hotter and much less dense, expanding envelope. This is more visible on the larger scale,
where a system of nested shocks is created by
this expansion against the collapsing gas within the main filament. The overall configuration
is that of a small dense core surrounded by expanding hot bubbles driven mostly by gas pressure and a 
non-negligible contribution from radiation pressure gradients. 

An interesting feature is that the core is colder than the expanding bubbles above 
the photosphere, as the temperature map conveys. This, in tandem with lower density above $R_{\rm ph}$, 
allows the radiation force to be 
more important at and above the photosphere. It also explains the driving forces behind the outflows.

By $t\sim 12.5$\,yr, the radiation force becomes comparable to the gas thermal pressure gradients 
(Fig.\,\ref{fig:FLD_accel}c), dramatically increasing the mass outflow rate. Figure\,\ref{fig:FLD_Rph} 
shows the evolution of the photosphere, which by this time becomes very extended, well outside the colder 
core. The core mass decreases sharply, as it is ``eaten away'' by the outflow. 

This process starts at around 
$t\sim 8$\,yr, when a strong outflow develops and extends up to and above $\sim 10^{-4}$\,pc, as shown in 
Figure\,\ref{fig:FLDproj2} at $t=9.5$\,yr. By $t\sim 15$\,yr, the core dissolves completely and the 
photosphere disappears.  

We have checked the existence of the photosphere using two independent methods outlined in
section\,\ref{sec:rad-transf}, by calculating  the 3-D shape of the photosphere using (1) the values 
of the limiter, $\lambda=1/3$, as a trace of the optically-thick region, and (2)  the optical 
depth by integrating along about 4,900 independent rays to $\tau=1$. 
Both methods have produced similarly shaped contours, with the latter contour lying at slightly larger
radii. The result of the first method is displayed in
Figure\,\ref{fig:FLD_Rph}. We have also introduced a 
spherically-averaged photospheric radius $R_{\rm ph}$, which we use in the associated discussion.
Figure\,\ref{fig:FLD_Lrad} displays the photospheric radius calculated using this method.

A second core forms nearby, separated by $\sim 3\times 10^{-4}$\,pc from the first core, and has an
initial mass of $\sim 0.01\,M_\odot$, but does not show any growth for a few years.
It starts to grow rapidly after $t\sim 25$\,yr. 
The new photosphere appears at $t\sim 30$\,yr and the core mass reaches $\sim 1\,M_\odot$ in about 
five years, exhibiting a growth rate of $\sim 0.2\,M_\odot\,{\rm yr^{-1}}$ (Fig.\,\ref{fig:FLD_Rph}).
By the end of the run, the central density of the second core, $\sim 3 \times 10^{-8}$ g cm$^{-3}$, 
had not yet reached the peak density of the first core, but it is still increasing with
time (Fig.\,\ref{fig:FLD_profile1}a).

Because of the perturbing action of the mass inflow, variations in $\rho$ and $T$, and the dependence of 
opacity on these parameters, the position of the photosphere, $R_{\rm ph}$, is erratic and it is far
from having spherical symmetry.
This is similar to the adiabatic run. Figure\,\ref{fig:FLD_Lrad} (top frame) provides the evolution
of the spherically-averaged $R_{\rm ph}$ with time. The photosphere for the FLD run is close to that 
of the adiabatic model initially, before the outflow develops. Within the photosphere,
however, the evolution differs significantly, e.g., in the importance of radiation pressure and rotation, 
and in the overall outcome. 

The central objects appear well resolved during the simulations. Their masses, $\sim 1\,M_\odot$,
are well above the local cell mass of $\sim 10^{-6}\,M_\odot$. The resolution limit is $\sim 3\times 
10^{-7}$\,pc.

For an isolated virialized system, the virial ratio is $X=2 E_{\rm kin}/|W|=1$, where 
$E_{\rm kin}$ is the total kinetic energy within the object, including the bulk and random 
motions, the radiation pressure is still not important, and $W$ is its gravitational energy. Since
cores obtained in our simulations are accreting
at a high rate kinetic and thermal energies as well, and experience mass loss, and we must also
include the relevant surface term in calculating $X$ \citep[e.g.,][]{lan80}.

For each of the cores formed, we calculate their virial ratios, $X(t)\equiv 
(2E_{\rm kin}-3P_{\rm ph}V)/|W|$, assuming their spherical symmetry, as a function of time. $X$ is 
calculated assuming the boundary of 
the object's ``surface'' lies at $R_{\rm ph}\sim 10^{-5}$\,pc. Here $P_{\rm ph}$ is the total pressure, 
i.e., the thermal and kinetic energies of the gas at $R_{\rm ph}$, and $V$ is the volume of the object. 
We account for the gas thermal energy in the accretion term because the radial velocity of the flow
is of the order of its sound speed. $3P_{\rm ph}V$ corresponds to the surface term in the Virial Theorem.

The contribution of the surface term, which consists
of the flow of the bulk kinetic and thermal energy of the gas should reduce the $X$ value below
unity. The sign of $P_{\rm ph}$ term depends on the relative importance of outflow and accretion averaged 
over the surface. It could be positive or negative, so the surface term could increase or decrease $X$.   
Indeed, this is what is observed --- $X$ varies below unity initially,
which delineates the unsteady contribution of the mass accretion flux. For the first core, $X$ becomes
larger than unity thereaqfter and steadily increases, reflecting the dissolution of the core.

\subsection{The Photosphere: radiation luminosity}
\label{sec:lumin}
 
Accreting mass flux carries a substantial kinetic energy because of the large $\dot M_{\rm acc}$.  What is 
the efficiency of converting this mechanical energy into radiation?

The kinetic energy of the accretion flux, measured at $R_{\rm ph}$, varies by about one decade within 
$L_{\rm acc}\sim 5\times 10^{37} - 5\times 10^{38}\,{\rm erg\,s^{-1}}$, and is of the order of the Eddington
luminosity, $\sim 10^{38}\,{\rm erg\,s^{-1}}$, for electron scattering opacity (Fig.\,\ref{fig:FLD_Lrad}b). 
Note that the Rosseland
mean opacity we use is of the order of the electron scattering opacity for these temperature and density
values. The range in $L_{\rm acc}$  is determined by
motion of the photospheric radius and temporal variation of the mass accretion rate and radial inflow velocity
(Figs.\,\ref{fig:FLD_profile1} and \ref{fig:FLD_profile2}).  The largest dip in $L_{\rm acc}$ is strongly
correlated with the dissolution of the first core, and the associated mass outflow in the region close to
$R_{\rm ph}$ (Fig.\,\ref{fig:FLD_Lrad}b). This process slows down the mass influx within the central 
$\sim 10^{-4}$\,pc. The influx is restored 10\,yr later, but it becomes much more noisy. 

The evolution of radiation luminosity, $L_{\rm rad}$, at some periods, correlates
with $L_{\rm acc}$, in other periods it anti-correlates (Fig.\,\ref{fig:FLD_Lrad}b). During the 
monotonic growth periods of both cores,
it clearly correlates. This behavior is disrupted by the powerful outflow which is associated with the
dissolution of the core. 

We have performed a Fourier analysis of the $L$ and $L_{\rm acc}$ curves in 
Figure\,\ref{fig:FLD_Lrad}b. The power spectrum of $L_{\rm acc}$ variability peaks around 
the characteristic timescale of $\sim 10$\,yr. It 
corresponds to the accretion timescale for a typical distance of ${\rm few} \times 10^{-5}$\,pc and the observed
inflow velocities of $\sim 3\,{\rm km\,s^{-1}}$. However, this timescale should be taken with caution,
as the simulation has been run only for about 40\,yrs, and so this timescale can be subject to 
temporal aliasing.
 
Additional and more rapid variability in $L_{\rm rad}$ is present at all times, but its amplitude increases 
following dissolution of the first 
core. The power spectrum also has a low peak at the  characteristic timescale of $\sim 0.12$\,yr. 
Typically, $L_{\rm rad}$ correlates with the accretion rate, but in some cases
the response in $L_{\rm rad}$ is either delayed or non-existent. During peaks of this variability, the radiation
luminosity can exceed the accretion power by a factor of a few, and $L_{\rm rad}$ can exceed 
$\sim 10^{39}\,{\rm erg\,s^{-1}}$. Clearly, energy can be stored within the photosphere, in either mechanical
or radiative form, and released suddenly.
 
\section{Discussion}
\label{sec:discuss}

 \begin{figure}
\centerline{
 \includegraphics[width=0.5\textwidth,angle=0] {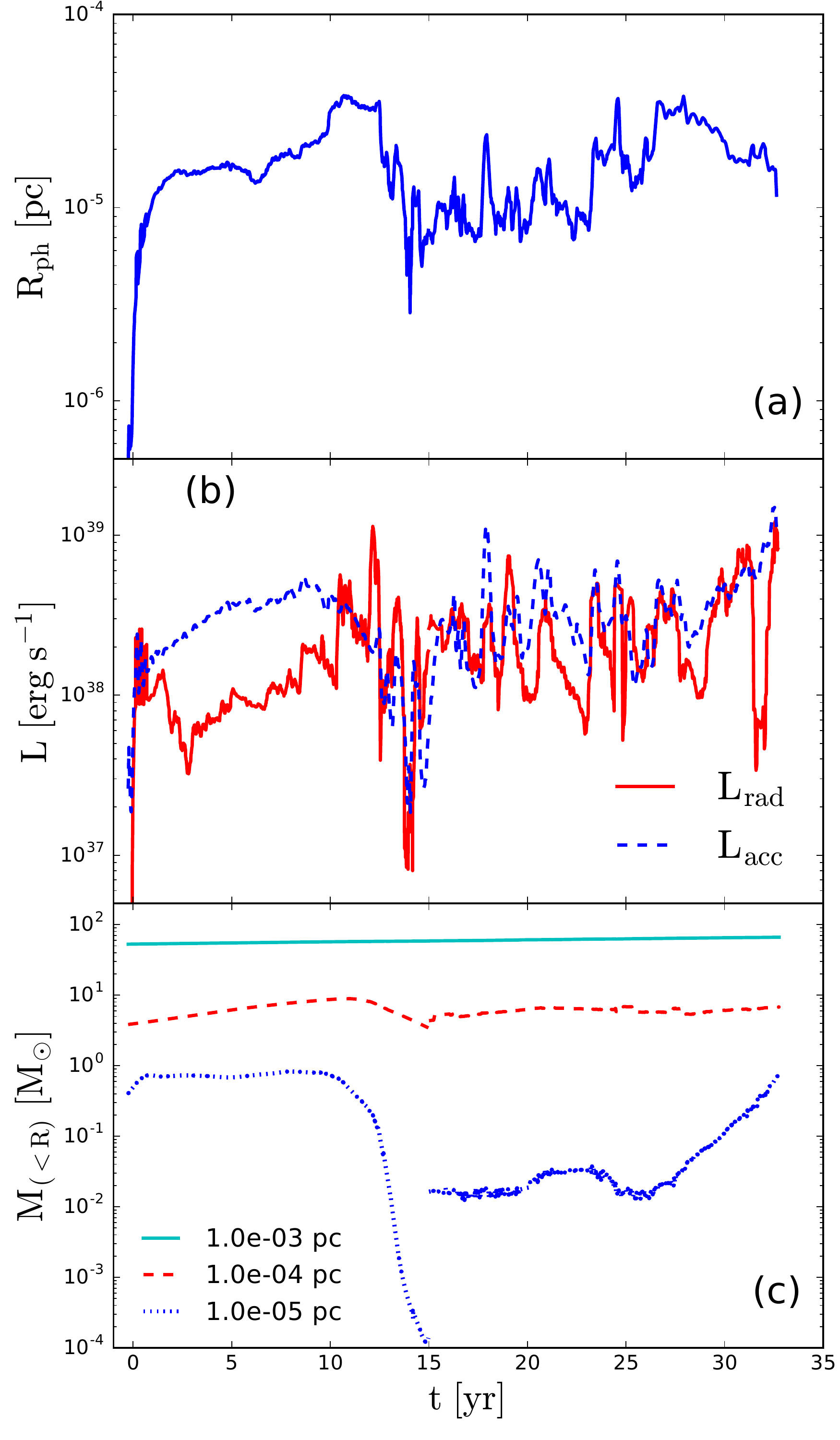}
}
\caption{{\it (a)} Evolution of the spherically-averaged photospheric radius, $R_{\rm ph}$, of the first core ($t\ltorder 15$\,yr) 
and the second core thereafter. {\it (b)} Evolution of radiation luminosity, $L_{\rm rad}$, and accretion 
(mechanical) luminosity, $L_{\rm acc}$, in the non-adiabatic model based on the photospheric radius shown. 
{\it (c)} Evolution of the enclosed mass within a fixed 
radius. The discontinuity in the dotted line reflects the dissolution of the first core at $t\sim 15$\,yr 
and the subsequent growth of a nearby core.
}
\label{fig:FLD_Lrad}
\end{figure}
 
 \begin{figure}
\center
 \includegraphics[width=0.5\textwidth,angle=0] {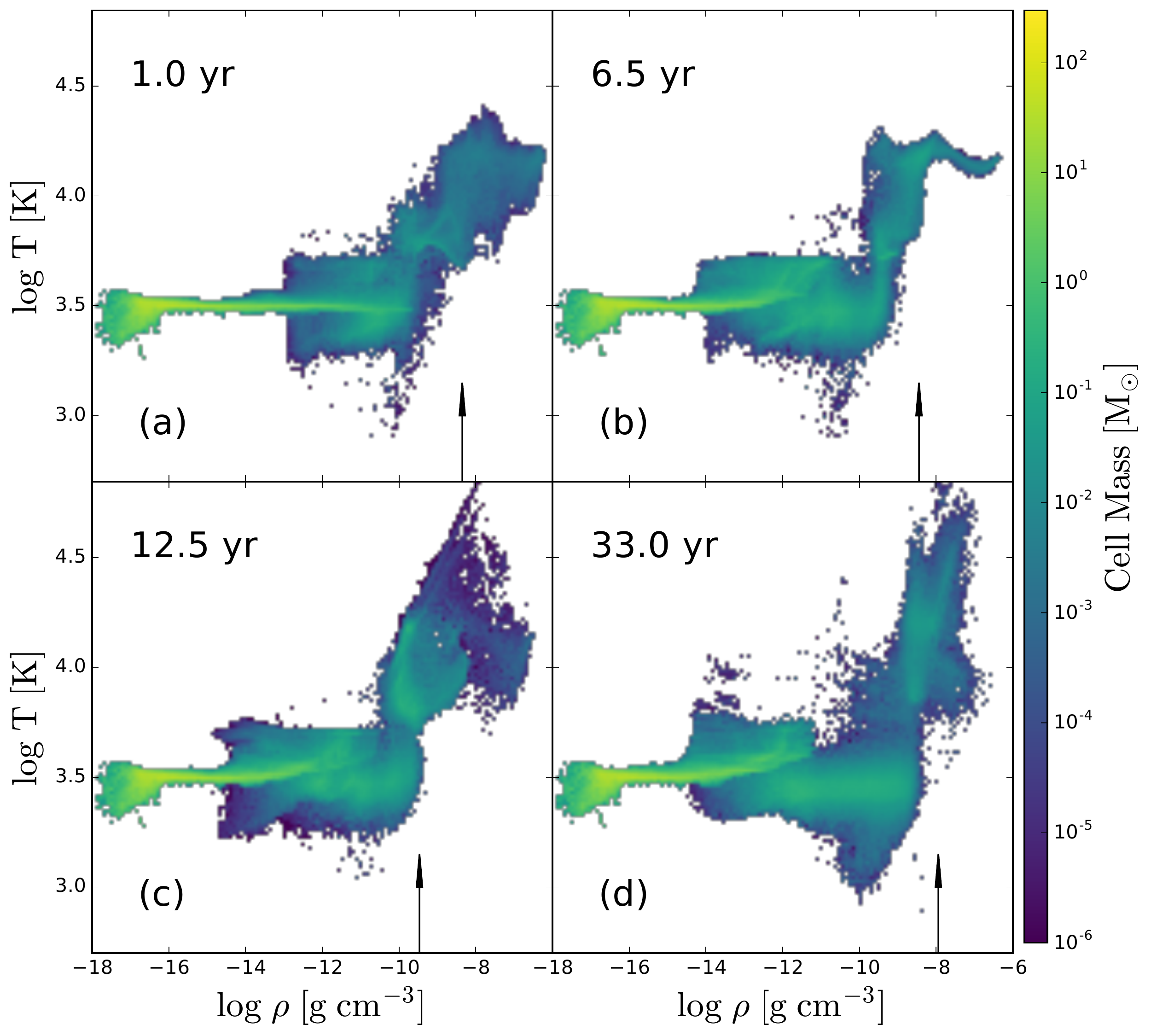}
\caption{Evolution of the non-adiabatic accretion flow with FLD: temperature versus gas density, at
$t\sim 1$\,yr, 6.5\,yr, 12.5\,yr and 32.7\,yr. The color palette shows the total mass of all grid cells
with the same density and temperature. The vertical arrows show the approximate position of the photosphere.
}
\label{fig:FLD_T_rho}
\end{figure}

We have followed direct baryonic collapse
within isolated DM halos. Inclusion of radiative transfer and the associated physics have
allowed us to reach spatial scales of $\sim 0.01$\,AU, or $\sim 10^{-7}$\,pc, for the
first time in a meaningful way.
The radiative transfer has been performed in the FLD approximation, and LTE
has been assumed for the optically-thick collapsing region. 

For comparison, we have run an adiabatic model, where the cooling rate has been exponentially
damped below the $\tau=1$ surface. We have tested the code by running a
number of models describing the evolution of a shock induced by a photoionization source at the
center of a hydrogen cloud. These models have been executed with FLD, and with and without LTE,
as detailed in the Appendix. Moreover, they have been compared
to published models in the literature, where analytical fits have been provided.
  
We find that the collapse proceeds in a filamentary way, and remains nearly isothermal in the
outer part, down to $\sim 10^{-5}$\,pc from the center. The gas is channeled  along
the filaments, with oblique shocks formed by the material when joining the filaments. Inside the
optically-thick region, a central object forms in response to the converging flow, and reaches a mass
of $\sim 1\,M_\odot$. Growing radiation and thermal pressure gradients within the object
exceed the gravitational acceleration, triggering a strong outflow, originating close to the
photospheric radius. The outflow has a bursting behavior, and 
drives expanding nested hot bubbles. The central core which forms deeper inside the photosphere
is close to dynamical equilibrium, but as the outflow `eats up' the core from outside in, the
core dissolves completely and the optical depth of the central $\sim 10^{-4}$\,pc hovers around unity.
Another core forms in its vicinity and grows rapidly, reaching $\sim 1\,M_\odot$.
The region inside the photosphere and its structure are well resolved in our simulations.  

The FLD model leads to the formation of an object which is supported mostly by gas thermal pressure,
and with some degree of rotation in the outer sub-photospheric layers. While the photosphere has a
complex elongated shape, the core of the
object is quasi-spherical. This is in a stark contrast with the adiabatic model,
where the central object is disky and of a convex shape, and is dominated by rotation.

The photosphere forms somewhat later, by $\sim 10^4$\,yr, in the FLD run --- a consequence of 
additional radiation force operating in the region. (Note, that initial conditions are identical for both 
runs.) If one compares both runs at $t\sim 33$\,yr, when the FLD model has been terminated, substantial 
differences point to diverging evolution. 

Specifically, the adiabatic model has a higher 
temperature in the central region, by a factor of 3, due to inability of the optically-thick flow to cool 
down. And the central mass accumulation is higher than in the FLD case, where a combination of radiation
force and thermal gas pressure gradients has driven a massive outflow.
These factors leads to a different radial profile of the specific angular momentum in the gas.
In particular, the ratio of the angular momentum to the maximally allowed value, is about
unity in the adiabatic case --- a clear sign of a rotational support --- whereas this ratio is smaller 
by a factor 
of a few in the FLD run. Consequently, the kinematics of the adiabatic flow differs from that of the FLD 
runs. Lastly, the adiabatic model shows fragmentation
on scales of $\sim 10^{-4}-10^{-3}$\,pc, while no fragmentation has been observed in the FLD runs.

We argue that the initial mass of the central objects, $M_0$,
can be understood in the context of a high accretion rate flow. For the object to be at least partially 
virialized, its sound crossing time should be faster than the characteristic timescale of its growth. 
The initial size of the object is, using the FLD run, $R_0\sim 10^{-5}$\,pc, 
and its gas temperature, $T_0\sim 10^4$\,K. Taking a typical mass accretion rate
in the central region (Figs.\,\ref{fig:adia_profile1} and \ref{fig:FLD_profile1}), $\dot M\sim 
0.1\,M_\odot\,{\rm yr^{-1}}$, we have

\begin{equation}
\label{eq:timescale}
\frac{R_0}{c_{\rm s}} < \frac{M_0}{\dot M},
\end{equation}
where $c_{\rm s}=1.3\times 10^6 (T/10^4\,{\rm K})^{1/2}\,{\rm cm\,s^{-1}}$ is the sound speed
in the gas. The smallest object in virial equilibrium under these conditions can be
estimated as $M_0\sim \dot M (R_0/c_{\rm s})\sim 0.1\,M_\odot\,(T/10^4\,{\rm K})^{-1/2}$. This 
result follows from the ability of an object to establish a partial equilibrium and to keep its
identity under strong mass accretion flow. It is not related to numerical issues. 

This means that the central object will be identified in the simulation at around this mass,
and is expected to be in very rough equilibrium only, with a mixture of thermal and radiation pressure
gradient, gravity, rotation and internal turbulence. 
The reason why much smaller objects cannot be identified lies in the fact that 
smaller objects will be buffered substantially by the inflow, the position of their
center of mass will be destabilized, and their shape will be completely arbitrary. This is not 
a semantic difficulty, but rather a condition for the object to separate itself from the dynamic inflow. 

The next question to be answered is related to the difference between the adiabatic and FLD models.
In a simplistic argument one can make a case that the adiabatic equation of state adequately
describes the behavior of the gas when an optical depth exceeds unity and the cooling declines
exponentially. Initial conditions for the adiabatic and FLD runs are identical and cannot explain
the different outcome. Besides, for gas evolution, initial conditions play secondary role, as the
system quickly forgets them. What is the source of the diverging evolution of these models?

The adiabatic equation of state presumes that the cooling is completely unimportant, and individual 
parcels of the gas do not exchange energy even in the presence of temperature gradients. This
requirement may be too restrictive. In a system that is not virialized, and basically consists of
streamers originating from strongly anisotropic inflow and is loosely bound, large temperature 
gradients build 
up. This can be seen from Figure\,\ref{fig:FLD_T_rho}b, which shows the dispersion in the gas temperature
around the mean, given by Figure\,\ref{fig:FLD_profile1}. The meaning of this is that the photons leak 
along large temperature gradients. The non-spherical shape of the photosphere assists in this process.
This effect is absent in the adiabatic flow. 

Next, we discuss the central mass accumulation over the simulation time. Even over the 35\,yr run time
since the formation of the photosphere in the FLD flow, about $10\,M_\odot$ are expected to be added
at the photospheric radius. Figures\,\ref{fig:FLD_profile1}d and \ref{fig:FLD_Lrad} do not show such
evolution on the scale of $R_{\rm ph}$. On the other hand, Figure\,\ref{fig:FLD_profile1}c confirms
that the high accretion rate peak moves to larger radii, outside $R_{\rm ph}$. The explanation lies
with the evolution in the presence of a strong outflow that acts against the mass accumulation 
inside the photosphere. Instead the gas accumulates inside $R\sim 10^{-3}$\,pc, as shown
in Figure\,\ref{fig:FLD_Lrad}a, which displays the amount of material inside this radius.

Models that quantify the amount of gas in the central region, and that ignore the feedback,
show the fast assembly of a massive object there \citep[][]{shlo16}. The current FLD run argues
against this conception. What does appear as important is that radiation feedback has an effective
distance beyond which it can be ignored. The FLD run puts this radius at $\sim 10^{-3}$\,pc,
where by the end of the simulation about $100\,M_\odot$ has been accumulated. This is about 200\,AU
--- the size of the Solar System. Within the typical star formation framework, this is probably
nothing outstanding, when, e.g., an O star forms. What is different here is the rate of accretion
which exceeds that of the star formation by an order of magnitude. Thus, one cannot argue that radiation
feedback will terminate accretion on the ``protostar'' and its growth.

The present state of the central region in the FLD run can be characterized as in a ``splash'' stage.
The gas accretion flow converges in the center and gravity is not capable of confining the resulting
random motions to within the photosphere. The radiation force at the photospheric radius is close to
the Eddington limit given the amount of mass in the region and the radiative luminosity 
(Figs.\,\ref{fig:FLD_profile1}d and \ref{fig:FLD_Lrad}b). Overall, such conditions are not encountered
in the star formation process, where both mass accretion rates and inflow velocities are dramatically 
lower, implying that the virialization process is much less violent. 

Some hints for the further evolution of the system can be inferred. The radiation-driven outflow is 
confined to
within $\sim 10^{-3}$\,pc, where the kinetic energy of the accretion can contain the kinetic energy
of the outflow. Once the outflow is stopped, the gas will have no pressure or rotational support and
must resume the collapse. We anticipate that additional outflow stages will follow but with 
progressively smaller amplitudes. But this does not mean that the system will virialize easily. 
 
A separate question is whether the above evolution leads to the formation of a single massive object,
e.g., an SMS, which will virialize and whose central temperature will exceed ${\rm few}\times 10^6$\,K,
enabling the proton-proton chain of thermonuclear reactions, and further stabilizing the
SMS. Our FLD runs, which appear to be more realistic than the adiabatic ones, have less rotational
support in the center, yet it is not negligible. Continuing accretion will bring fresh material
with increasing angular momentum. The reason for this is that low-$J$ gas is naturally accreted
first, and the subsequent accretion will increase its $J$. Because of a large accretion rate, this
can lead to a spinup of the object, non-axisymmetric instabilities, and a resumption of the central 
runaway, similarly to the scenario that happened at $\sim 1$\,pc in the earlier stage.
 
The difference in the evolution between the adiabatic and FLD models emphasizes the importance
of the proper treatment of radiative transfer in the optically-thick phase of gravitational collapse.
This requires a 7 dimensional phase space to obtain the radiative intensity, 
which is impossible to achieve at present even numerically. Both Monte-Carlo and direct discretization 
methods require too many computational resources. Limiting calculations to radiative flux and energy
allows one to take the angular moments of the equations of radiation hydrodynamics. 
Examples of such low-order closures are FLD, discussed in \S2.1, and the $M_1$ closure 
\citep[e.g.,][]{lev84,jan92}. Their deficiency lies in an inadequate treatment of the Eddington
tensor, which is symmetrized about the direction of the flux. In certain
cases, this crude approximation can fail \citep[e.g.,][]{jia12}.

Both FLD and $M_1$ can be applied in the optically-thin and thick regions. Potentially, the FLD 
method can lead to errors, as it has difficulty to capturing the shadow formed even by one beam 
\citep[e.g.,][]{gonza07}, while the $M_1$ method cannot propagate two beams correctly, having difficulty 
following the radiation field in complex geometries \cite[e.g.,][]{mckin14}.

The weak point of both FLD and $M_1$ algorithms --- their difficulty in handling the transition between 
optically-thick and thin regions --- 
can be supplemented by the ray-tracing method. This approach was implemented in the PLUTO grid code 
for a spherical polar grid \citep[e.g.,][]{kui10}. As ray-tracing is a solution to the radiative transfer
equations, FLD is an approximation, there is a clear advantage in combining both methods 
\citep[e.g.,][]{kla14}. This means, using direct ray-tracing in optically-thin regions, where 
scattering can be ignored, while implementing FLD in optically-thick regions, where diffusion 
dominates. 

An algorithm that is based on the direct solution of the radiative transfer equations and that does not 
invoke a diffusion approximation has been
proposed for the MHD code Athena \citep{jia12}. The hierarchy of moment equations has been closed
using a variable Eddington tensor, whose components have been calculated using the method of
short characteristics, still computationally expensive. Further improvements must follow along these 
lines.

\section{Conclusions}
\label{sec:conc}

We have simulated the radiative transfer in gravitationally collapsing primordial gas within
isolated DM halos, so-called direct collapse. Models in the cosmological framework are dealt with
in an associated publication \citep[][]{arda17}. We focus on the optically-thick part of the collapse, 
initially at radii below
$\sim 10^{-6}$\,pc, 0.1\,AU, where the photosphere of the central object has formed. The radiative
transfer was performed in the flux-limited diffusion (FLD) approximation, using  
a modified version of the Enzo-2.4 AMR code, and LTE conditions were assumed. For comparison,
we have run adiabatic models, and additional testing of the FLD module is shown in the Appendix.

We find that the collapse is dominated by filamentary structure modified by rotation, down to the 
photospheric scale.
The central object that forms within the photospheric radius grows to $\sim 1\,M_\odot$,
and is supported mainly by thermal gas pressure gradients with the addition of rotation. The 
evolution of this
object is heavily perturbed by the penetrating accretion flow that peaked at 
$\sim 0.5\,{\rm M_\odot\,yr^{-1}}$, growing temperature and increasing
radiation pressure. The photospheric luminosity is close to the Eddington limit. This leads to the 
development of an anisotropic outflow driven by radiation force,
which disrupts the central object and dissolves it, driving a series of expanding hot bubbles interacting 
with the accretion flow. 

The dissolution of the core leads to the formation of another core
nearby, which grows efficiently and shortly reaches $\sim 1\,M_\odot$. With the formation of this object 
the central temperature starts to grow, sharply decreasing the timestep.
At this point, the enclosed mass within the central $10^{-3}$\,pc, is about $100\,M_\odot$, with about 
$3\times 10^3\,M_\odot$ within the central 0.1\,pc.

This mass accumulation agrees with that of the adiabatic run, but its kinematics is substantially different. 
The adiabatic run forms a geometrically-thick disk, supported mainly by rotation with an admixture of 
thermal gas pressure. Outside this disk a number of fragments form which show a tendency to merge with
the central convex-shaped disk. This fragmentation is observed on scales 
between $\sim 10^{-4}$\,pc and $10^{-3}$\,pc, and temporarily disrupts the growth of the central object. 
This object is in contrast with quasi-spherical shapes of the forming cores in the FLD case. 

In both cases, the photospheric shapes are very irregular, which allows the radiation to diffuse out
of the central region. This explains the major difference between the adiabatic and the FLD runs, and
reveals the inapplicability of the adiabatic approximation to the growth of the central core in direct 
collapse.

We find that the typical radiation luminosity from the photosphere of each of the cores formed lies in the 
range $\sim {\rm few} \times 10^{37}
-{\rm few} \times 10^{38}\,{\rm erg\,s^{-1}}$ over much of the run time. This is of order the 
Eddington luminosity for such an object. Fourier analysis shows that this luminosity varies
on two characteristic timescales: a long one, which is associated with the variable accretion timescale, 
and $\sim 0.1$\,yr,
which originates in the radiative diffusion timescale within the photosphere. The latter variability is 
characterized by a large amplitude which exceeds $10^{39}\,{\rm erg\,s^{-1}}$.

This study reveals that models accounting for radiative transfer in the collapsing gas display a different
evolution than models with an adiabatic equation of state, at least during early stages of core formation. 
The main reason for these differences
is that the radiation is capable of diffusing out due to the anisotropy in density and temperature,
and the resulting decrease in opacity in various directions. This effect vanishes in the 1-D case and requires 
a multi-dimentional
treatment. The underlying gas dynamics changes as a result, leading to massive outflows from forming cores.
It modifies the angular momentum transfer, and the flow avoids fragmentation in the optically-thick regime,
which prevails in the adiabatic case.

\section*{Acknowledgments}
We thank the Enzo and yt support team for help. All analysis has been conducted using yt
\citep{yt}, http://yt-project.org/.
We thank Daniel Reynolds for help with the FLD solver, and Kazuyuki Omukai for providing the updated
opacities for a comparison. Discussions with Pengfei Chen, Michael Norman, Kazuyuki 
Omukai, Daniel Reynolds, and Kengo Tomida are 
gratefully acknowledged. This work has been partially supported by the Hubble Theory grant 
HST-AR-14584, and by JSPS KAKENHI grant 16H02163 (to I.S.). 
I.S. and K.N. are grateful for a generous support from the International Joint Research 
Promotion Program at Osaka University. JHW acknowledges support from NSF grant AST-1614333, Hubble Theory
grants HST-AR-13895 and HST-AR-14326, and NASA grant NNX-17AG23G. MB acknowledges NASA ATP grants 
NNX14AB37G and NNX17AK55G and NSF grant AST-1411879. The STScI is 
operated by the AURA, Inc., under NASA contract NAS5-26555. Numerical simulations have been
performed on XC30 at the Center for Computational Astrophysics, National Astronomical Observatory of Japan,
on the KDK computer system at Research Institute for Sustainable Humanosphere, Kyoto University,
on VCC at the Cybermedia Center at Osaka University, as well as on the DLX Cluster of the University of 
Kentucky.

\appendix

\section{Expansion of an H\,II region around a point source of radiation: the role of the radiation force}
\label{sec:app}

In order to test our version of Enzo, we compare the analytical and numerical solutions for an envelope expanding 
away from a point source of radiation and accelerated by thermal pressure gradients from photoionization and by 
radiation force \citep[e.g.,][]{wise12,rosd15}.
The analytical solution is based on momentum conservation in the swept gas around the central ionizing source
with luminosity $L$, neglecting 
the terms associated with gravity, heating and cooling, so that $\dot {p}= L/c$, where $p$ is  
momentum, and $c$ is the speed of the light. The resulting  radial 
position $r(t)$ of the expanding H\,II  front for an initially uniform gas $\rho_0$ is 
given by:
\begin{equation} \label {eq1}
r(t)=(R_{\rm s}^4+2At^2)^{1/4},
\end{equation}
where $R_{\rm s}=(3\dot {N}_{\rm \gamma}/4\pi n_{0}^2\alpha_{\rm B})^{1/3}$ is the Str\"{o}mgren sphere radius, 
$A=3L/4\pi\rho_{0}c$, $\alpha_{\rm B}=2.5\times10^{-13}\,{\rm cm^3\,s^{-1}}$ is the case B recombination rate 
at $T=10^4$\,K, $n_0$ is the hydrogen number density, and $\dot {N}_{\rm \gamma}$ is the 
rate of emitted photons per 
second from the source. An additional effect is due to the gas thermal pressure which 
results from photoionisation heating. In the absence of radiation force, the 
H\,II front expands due to the photoionisation heating as \citep{spi78}:
\begin{equation} \label {eq2}
r(t)=R_{\rm s}\left(1+\frac{7c_{\rm s}t}{4R_{\rm s}}\right)^{4/7}
\end{equation}
where $c_{\rm s}$ is the sound speed in the ionized gas.

To study the role of the radiation force, we set up a cubic box and place a point source 
$L=10^{6}L_\odot$ at the center of the box. The simulation box is resolved with $128^{3}$ cells. The point 
source emits ionizing photons in the energy band 13.6 -- 24.6\,eV into an initially uniform neutral pure 
hydrogen gas at a temperature of $T_0=10^{3}$ K. The 
tests are performed for three different initial gas number densities, namely, $10^{5}$, $10^{7}$, and 
$10^{9} {\ \rm cm^{-3}}$. For each number density, the simulation is performed with and without the 
radiation force, while the photoionization heating is present in both cases. The box size and run time 
for each test are summarized in the Table\,\ref{dsh}. For these tests we assume non-LTE conditions,
which means that we solve for the H-chemistry, do not assume Planckian emissivity, and calculate
emission versus absorption in this energy bin.
\begin{table}
\begin{center}
\caption{Simulations setup for expanding H\,II region.\label{dsh}}
\begin{tabular}{ |p{2cm}||p{1.5cm}||p{1.5cm}||p{1.5cm}|  }
 \hline
Parameter&  Test\,I& Test\,II&Test\,III\\
 \hline
$L$     &  $10^{6}L_\odot$& $10^{6}L_\odot$&$10^{6}L_\odot$\\
$n_{\rm H} $\    $({\rm cm}^{-3})$&  $10^{5}$& $10^{7}$&$10^{9}$\\
$L_{\rm box}$\  (pc)&  2& 0.2&0.02\\
$t_{\rm run}$\ (Myr)&  $1.2 \times 10^{-2}$& $2.1 \times 10^{-3}$&$2.7\times 10^{-4}$\\
 \hline
\end{tabular}
\end{center}
\end{table}

\begin{figure}
\begin{center}
\includegraphics[scale=0.45]{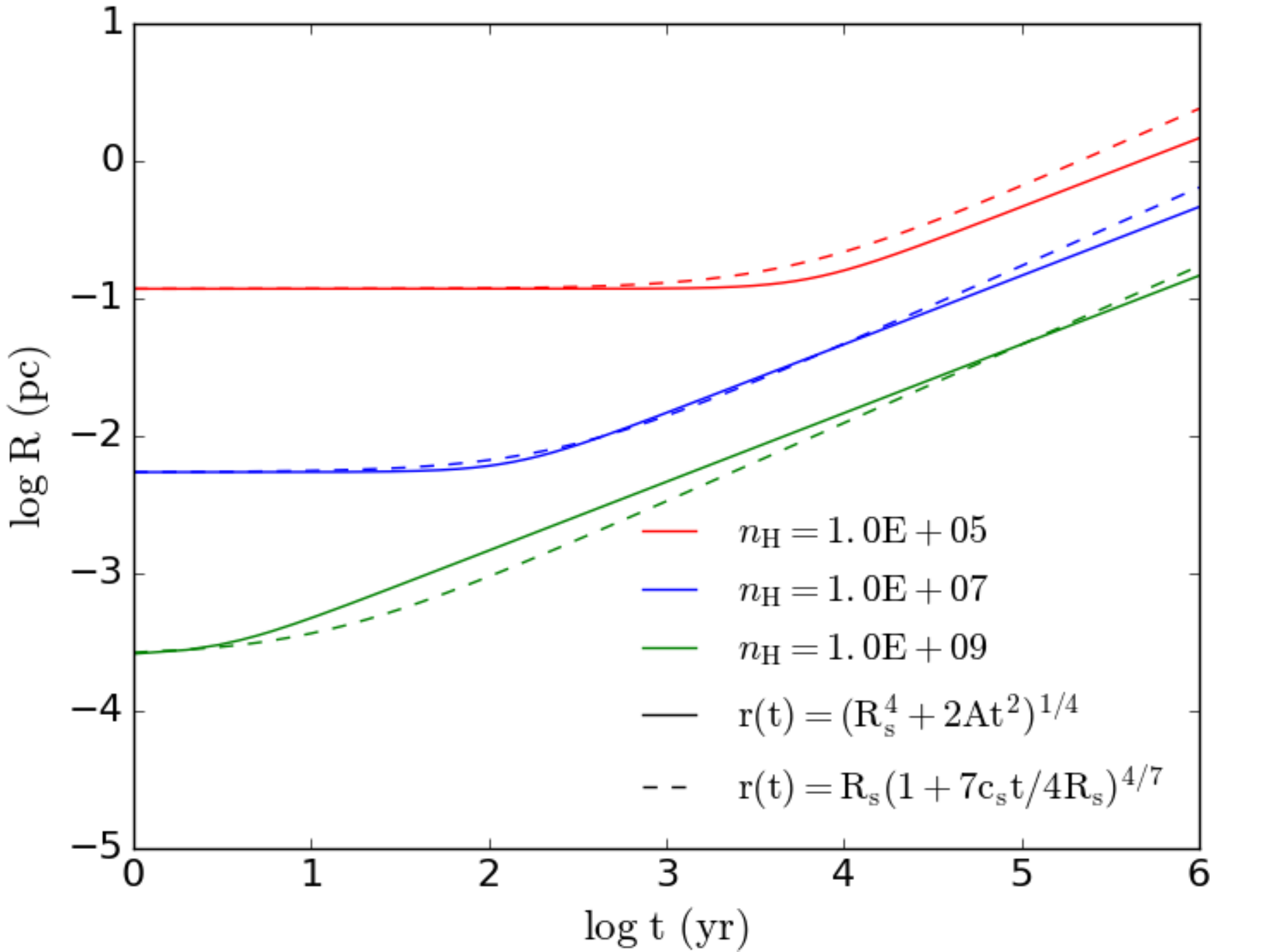}
\caption{Radius of the expanding shell calculated based on radiation force (Eq.\,\ref{eq1}, solid lines), 
and on the effect of the photoionisation heating 
(Eq.\,\ref{eq2}, dashed lines), for different hydrogen number density of $n_{\rm H}=10^{5}$, $10^{7}$, and 
$10^{9}\,{\rm cm^{-3}}$. For all cases, the point source luminosity is $L=10^{6}\,L_\odot$.}
\label{ana}
\end{center}
\end{figure}

Figure\,\ref{ana} shows the H\,II front expansion driven by a direct momentum absorption 
from the ionizing source (Eq.\,\ref {eq1}), or as a result of the photoionisation heating only (Eq.\,\ref {eq2}).  
It clearly shows that radiation force has a trivial contribution at the lowest density $10^{5}\,{\rm cm^{-3}}$, 
and the expansion is controlled by photoionisation heating (Eq.\,\ref{eq2}). As the density increases, 
at first, the contribution of radiation force in the expansion exceeds the photoionisation heating (see 
the green line in Figure\,\ref{ana} for time $\leq 10^4$ yr), and the photoionisation heating dominates 
the process afterwards (see Figure\,\ref{ana} for time $\geq 10^5$ yr).

\begin{figure}
\begin{center}
\includegraphics[scale=0.35]{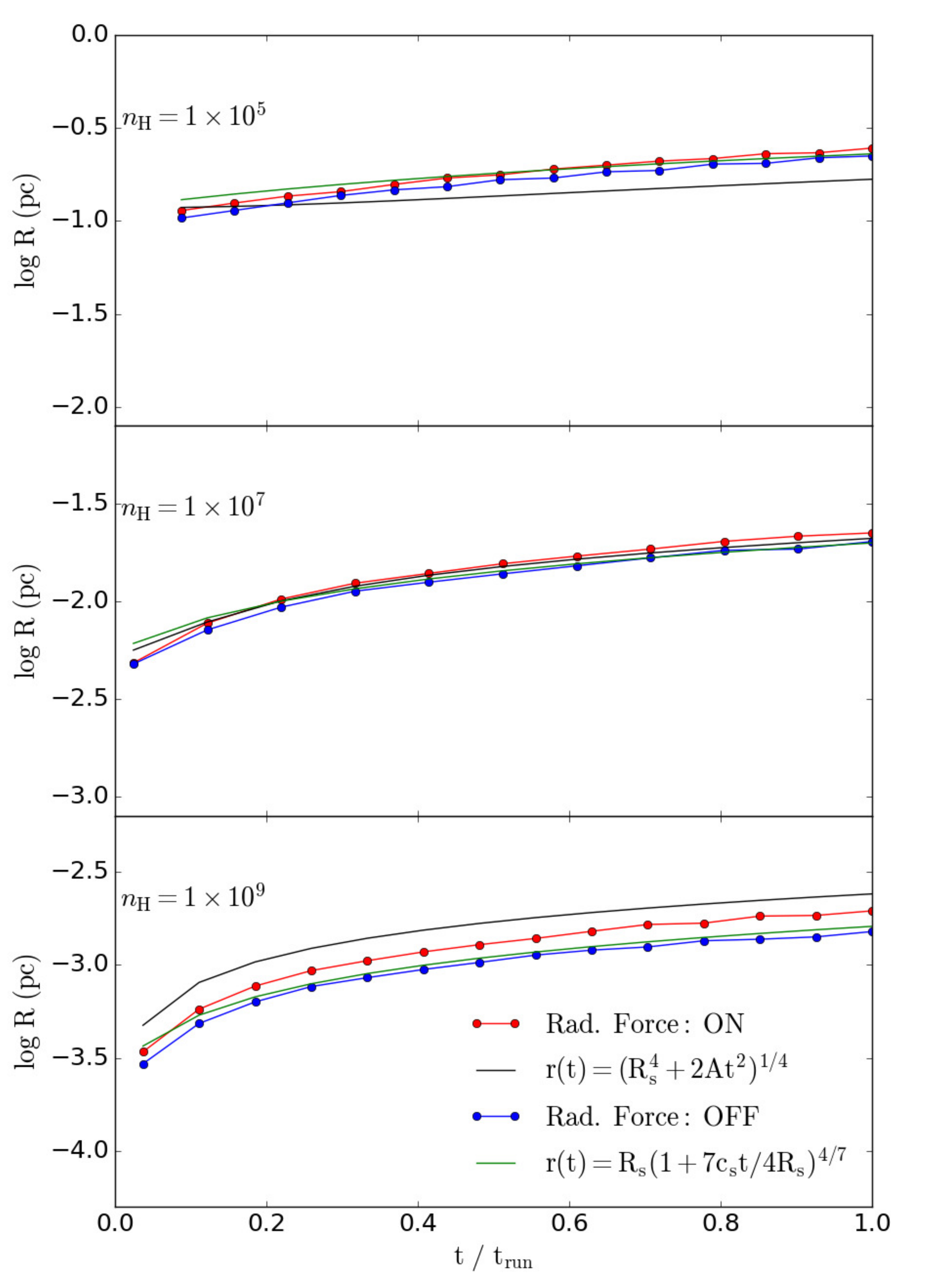}
\caption{Radius of the expanding H\,II region versus time for the numerical simulation including the 
radiation force (red circle) and without radiation force (blue circle). The analytical evolution of the radius 
due to the radiation force (black dashed line, Eq.\,\ref{eq1}) and due to the photoionisation 
heating (green dashed line, Eq.\,\ref{eq2}) are also provided.}
\label{num1}
\end{center}
\end{figure}

For the performed tests, the H\,II front radius is compared with  Eq.\,\ref{eq1}, for the cases 
when the radiation force has the dominant effect, and compared  with Eq.\,\ref{eq2}, when the effect 
of photoionisation heating is dominant. Shown in Figure\,\ref{num1} is the radial evolution of the expanding 
H\,II region for different densities. The radius is estimated to be located where the neutral fraction 
$x_{\rm H\,I} = 0.5$. As discussed before, the radiation force has a negligible effect for the case of 
$n_{\rm H}=10^{5}\,{\rm cm^{-3}}$. Therefore, the 
H\,II front radius is mainly determined by the photoionization,  and the simulations with and 
without the radiation force yield quite similar results (e.g., see panel $a$). As the density increases, the 
radiation force becomes more important. For density $n_{\rm H}=10^{7}\,{\rm cm^{-3}}$, panel $b$ shows that the  
radiation force has small additional contribution to the H\,II front expansion, which agrees with the 
analytical solution. 
A more significant effect of the radiation force was found for a density of $n_{\rm H}=10^{9}\,{\rm cm^{-3}}$
(panel $c$), where the expanding H\,II region is governed by the radiation force (see the red circles), 
and its front radius is well approximated by Eq.\,\ref{eq1}, black line. The correspondence between the 
analytical and numerical solutions is very good.

\begin{figure}
\begin{center}
\includegraphics[scale=0.37]{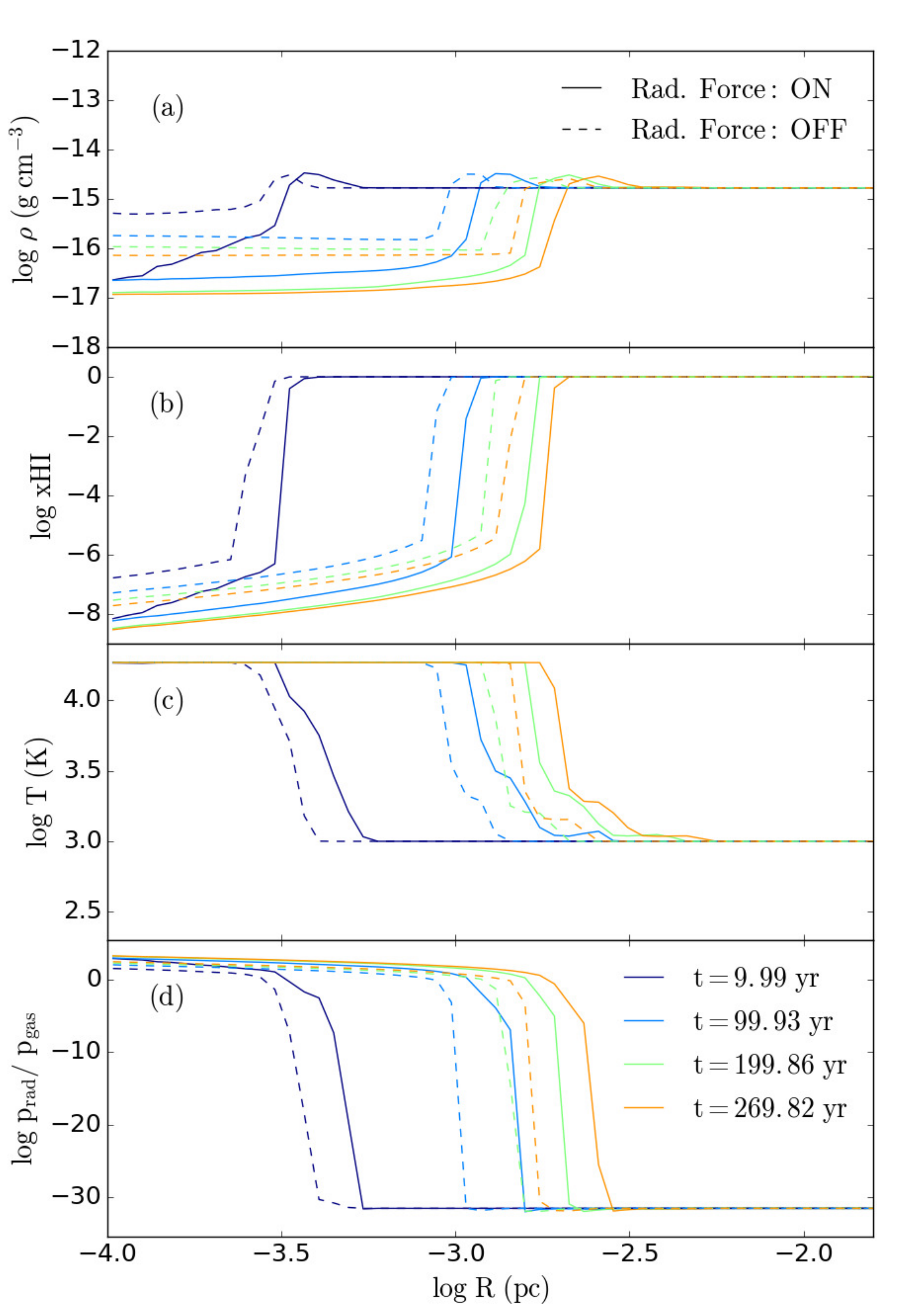}
\caption{Radial profiles of (a) the gas density, (b) the neutral fraction, (c) the temperature, and 
(d) the ratio of 
the radiation pressure to gas pressure given at different times for the initial hydrogen number density 
is $n_{\rm H}=10^{9}\,{\rm cm^{-3}}$. Runs with only photoionisation heating are represented by dashed curves, 
while runs that include a radiation force from the ionizing photons are given by solid curves.}
\label{num2}
\end{center}
\end{figure}

Shown in Figure\,\ref{num2} are radial profiles of (a) the gas density, (b) the neutral fraction, 
(c) the temperature, 
and (d) the ratio of radiation-to-thermal pressure for an initial number density of $10^{9}\,{\rm cm^{-3}}$. 
The profiles of 
density and neutral fraction in each figure (panels $a$ and $b$) clearly demonstrate the expansion of the
H\,II region. For this case, the radiation force is dominant. The gas density and hence 
the neutral fraction substantially decrease in the H\,II region (by about two orders of magnitude), 
and the ratio of radiation-to-thermal pressure increases by almost two orders of magnitude. 
Therefore, the resulting expansion of the bubble is mainly driven by a direct radiation force. 

There are two important issues to point out. Firstly, for the case of a dominant radiation pressure, the 
bubble expansion will be stalled at a radius $R_{\rm 1}$, where the outwards radiation pressure from the point 
source is equal to the thermal pressure from outside the bubble, $L/4\pi R_1^2 c=n_{\rm H} k_{\rm B}T_0$. 
Secondly, in regards to dominant photoionisation heating, the expansion stops when 
$n_{\rm b} T_{\rm b}=n_{\rm H} T_0$, where $n_{\rm b}$ and $T_{\rm b}$ are gas density and temperature in the
bubble. The radius of the bubble in this case can be estimated from $\dot {N}_{\rm \gamma}=4/3\pi 
R_2^3\alpha_{\rm B}n_{\rm b}^2$. Within this radius, the ionizing luminosity of the point source provides an equal 
rate of photoionizations to the recombination rate within the bubble. 

As one can see, the terminal radius of the bubble cannot be determined using Eq.\,\ref{eq1} and Eq.\,\ref{eq2}
and happens to lie well outside our calculation domain. \citet{rosd15} presented the expanding H\,II 
region runs with the RAMSES--RT code. In these runs, the maximal bubble radii have been reached and
agreed well with $R_1$ and $R_2$ for the dominant radiation pressure and photoionisation heating, respectively.


\end{document}